\documentclass{article}
\usepackage[utf8]{inputenc}

\usepackage{algorithm}
\usepackage{algpseudocode}
\usepackage{amsfonts}
\usepackage{amsmath}
\usepackage{amsopn}
\usepackage{amssymb}
\usepackage{array}
\usepackage{authblk}
\usepackage{bbold}
\usepackage[makeroom]{cancel}
\usepackage{comment}
\usepackage{dutchcal}
\usepackage{enumitem}
\usepackage{epstopdf}
\usepackage{float}
\usepackage{graphicx}
\usepackage{geometry}
\usepackage{mathtools}
\usepackage[numbers]{natbib}
\usepackage{pdfpages}
\usepackage{pgfgantt}
\usepackage{siunitx}
\usepackage{soul}
\usepackage{stix}
\usepackage[UKenglish]{babel}
\usepackage{url}
\usepackage{wrapfig}
\usepackage{xcolor}

\usepackage[colorlinks,citecolor=blue,bookmarks=false,linkcolor=blue,urlcolor=blue]{hyperref}

\setlist[enumerate]{leftmargin=.5in}
\setlist[itemize]{leftmargin=.5in}

\DeclareMathOperator*{\Var}{\mathrm{Var}}
\DeclareMathOperator*{\Cov}{\mathrm{Cov}}
\DeclareMathOperator*{\E}{\mathbb{E}}
\DeclareMathOperator*{\MSE}{\text{MSE}}
\DeclareMathOperator*{\sgn}{\text{sgn}}
\DeclareMathOperator*{\Epsilon}{\mathcal{E}}
\DeclareMathOperator*{\bigO}{\mathcal{O}}
\DeclareMathOperator*{\smalls}{\mathbcal{s}}
\newcommand{\mvec}[1]{\boldsymbol{#1}}  

\newcommand{\surr}[1]{f^{(#1)}}  
\newcommand{\ntr}[1]{^{(#1)}}  
\DeclareMathOperator*{\s2}{{\sigma^2}}  

\newcommand{\red}[1]{#1}

\newtheorem{theorem}{Theorem}[section]
\newtheorem{proof}{Proof}[section]
\newtheorem{lemma}{Lemma}[section]
\newtheorem{remark}{Remark}[section]
\newtheorem{assumption}{Assumption}[section]      

 \title{Lasso Monte Carlo, \red{a Variation on Multi Fidelity Methods} for High Dimensional Uncertainty Quantification}
\author[1,2]{Arnau Alb\`a}
\author[1]{Romana Boiger}
\author[1]{Dimitri Rochman}
\author[*,1]{Andreas Adelmann} 
\affil[1]{Paul Scherrer Institut, Forschungstrasse 111, 5232 Villigen, Switzerland}
\affil[2]{ETH Z\"urich, R\"amistrasse 101, 8092 Zurich
 Switzerland}
\affil[*]{andreas.adelmann@psi.ch}
\date{\today}
 
\begin{document}
\baselineskip=16pt
\parindent=15pt
\parskip=4pt

\maketitle

\begin{abstract}
    Uncertainty quantification (UQ) is an active area of research, and an essential technique used in all fields of science and engineering. The most common methods for UQ are Monte Carlo and surrogate-modelling. The former method is dimensionality independent but has slow convergence, while the latter method has been shown to yield large computational speedups with respect to Monte Carlo. However, surrogate models suffer from the so-called \textit{curse of dimensionality}, and become costly to train for high-dimensional problems, where UQ might become computationally prohibitive. In this paper we present a new technique, Lasso Monte Carlo (LMC), which \red{combines a Lasso surrogate model with the multifidelity} Monte Carlo technique, in order to perform UQ in high-dimensional settings, at a reduced computational cost. We provide mathematical guarantees for the unbiasedness of the method, and show that LMC \red{can be more accurate} than simple Monte Carlo. The theory is numerically tested with benchmarks on toy problems, as well as on a real example of UQ from the field of nuclear engineering. In all presented examples LMC \red{is more accurate} than simple Monte Carlo \red{and other multifidelity methods. Thanks to LMC,} computational costs are reduced by more than a factor of 5 with respect to simple MC, in relevant cases.
\end{abstract}

\section{Introduction}

Uncertainty Quantification (UQ) aims to calculate the effect of unknown or uncertain system parameters on the outcome of an experiment or computation. It is an active area of research, and an essential tool to test the robustness and accuracy of methods used in many domains of science and engineering, such as risk assessment in civil engineering \cite{nagel_principal_2020}, design and optimisation of particle accelerators \cite{adelmann_nonintrusive_2019, frey_global_2021}, weather prediction \cite{allen_quantifying_2000}, medical physics \cite{rizzoglio_uncertainty_2020}, and nuclear engineering \cite{leray_nuclear_2016, rochman_nuclear_2016, rochman_uncertainties_2018}.

The UQ process can be described as follows: let $f(\mvec x)$ be a deterministic function which represents the numerical experiment

\begin{equation*}
    f\colon \begin{array}[t]{ >{\displaystyle}r >{{}}c<{{}}  >{\displaystyle}l } 
          \mathbb{R}^{d} &\to& \mathbb{R}^m \\ 
          \mvec x &\mapsto& f(\mvec x)\,, 
         \end{array}
\end{equation*}

with input and output dimensions of size $d$ and $m$, respectively. Let \red{$\mvec x_0=(x_1,x_2,...,x_d)$} be an input vector, with an associated uncertainty. The uncertainty can be modelled by letting the input be random variable $X$, centred around $\mvec x_0$, such that $\E[X] = \mvec x_0$. A common approach is to model the input with a multivariate normal distribution $X\sim\mathcal{N}(\mvec x_0,\Sigma)$, where $\Sigma$ is a known covariance matrix. The aim of UQ, and more specifically \textit{response variability methods}, is to estimate the mean $\mu$ and variance $\sigma^2$ of the output distribution $f(X)$ which is then written as $f(\mvec x_0) = \mu\pm\sigma$. Without loss of generality, in the rest of the paper it is assumed that the output is one-dimensional, i.e. $m=1$.

When $f$ is a black-box function one has to rely on non-intrusive UQ methods, such as Monte Carlo (henceforth referred to as simple MC) \cite{helton_survey_2006, arnst_overview_2014} or surrogate modelling \cite{sudret_surrogate_2017, owen_comparison_2017}.

With simple MC, $N$ \red{independent and identically distributed} (i.i.d). multivariate random variables $X_1,X_2,...,X_N$ are sampled to obtain a set of input vectors $\mvec x_1, \mvec x_2, ..., \mvec x_N$. Then $f$ is evaluated at each input to obtain a set of outputs $f(\mvec x_1), f(\mvec x_2), ..., f(\mvec x_N)$, from which the sample mean and sample variance are calculated. Monte Carlo methods are known to converge as $\bigO(N^{-{\frac{1}{2}}})$ which, if $f$ is expensive to evaluate, can make such methods computationally expensive or even prohibitive.

\red{The slow convergence of the errors with simple MC can in some cases be bypassed by using surrogate models \cite{sudret_surrogate_2017}. With this approach, the $N$ input-output samples of $f$ are used to train a surrogate model $\surr{N}$. Then $\surr{N}$ is evaluated $M$ times, and the outputs are used to estimate the mean and variance. The advantage of this method is that $\surr{N}$ is computationally cheaper than $f$, and evaluating it $M$ times with $M\gg N$ has negligible runtime. The bottleneck of this method is in obtaining the $N$ samples for the training set. If $f$ is low-dimensional such that $N\gg d$, the surrogate model is likely to have a small bias, however in high-dimensional cases with $N<d$ the surrogate will be biased (see the \textit{curse of dimensionality} \cite{beyer_when_1999, koppen_curse_2000}), making also the mean and variance estimations biased. The specific case where $N<d$, and where increasing $N$ is not possible or computationally expensive, is the main focus of this paper. The aim is to find a method more accurate than simple MC for a fixed computational budget $N$.}


\red{
In this regard, multifidelity Monte Carlo (MFMC) \cite{peherstorfer_survey_2018,Peherstorfer2016,peherstorfer_convergence_2018, sharifnia_multilevel_2022} offers a promising approach. Multifidelity methods combine low and high fidelity models to accelerate the solution of an outer-loop application, like uncertainty quantification, sensitivity analysis or optimisation. The low fidelity models, that are either simplified models, projection-based models or data-fit surrogates, are used for speed up, while the high fidelity models are kept in the loop for accuracy and/or convergence. Thus MFMC offers a framework to combine a biased surrogate model $\surr{n}$ with the high fidelity model $f$, in such a manner that unbiased estimates of the mean and variance are computed. The crucial point with MFMC is then, for a given number of high fidelity evaluations $N$, optimising the trade-off between how many of them are used in the multifidelity estimators and how many for training the surrogate.  This optimisation problem has been recently addressed in \cite{peherstorfer_adaptive_2019}. Nevertheless, there is no guarantee that, for a given $N$, the MFMC estimates will be more accurate than simple MC, especially in high-dimensional cases with $N<d$, where $\surr{n}$ is likely to have a large bias.}

To address the challenges presented by high-dimensional UQ in existing approaches, we introduce the Lasso Monte Carlo (LMC) method, a variation on MFMC. 
With LMC we propose a new data management strategy in which the high fidelity samples are reused several times both to train multiple surrogates and in the multifidelity estimators. The resulting algorithm is such that the estimations are guaranteed to be equally or more accurate than simple MC and MFMC, under certain assumptions. This new approach can be viewed as a variance reduction technique, based on a two-level version of MFMC, and Lasso regression (least absolute shrinkage and selection operator), simultaneously performing regression analysis, variable selection and regularization \cite{tibshirani_regression_1996}.

\red{It is worth mentioning the relation between multifidelity, multilevel, and multi-indexing methods: Multilevel Monte Carlo (MLMC) \cite{giles_multilevel_2008} is a special case within the broader framework of multifidelity methods, in which typically a hierarchy of low fidelity models is derived by varying a parameter (e.g. mesh width). A generalization of MLMC is introduced in \cite{haji-ali_multi-index_2016}, so called multi-index Monte Carlo methods, which use multidimensional levels and high-order mixed differences to reduce the computational cost and the variance of the resulting estimator. }


The remainder of the paper is organised as follows: we start with a review of MFMC for the estimation of central moments, and in particular we derive the expressions for the two-level estimators. \red{This is followed by a discussion on the trade-off between accuracy and computational costs of MFMC.} The new method LMC is then introduced, and we prove that it is equally or \red{more accurate} than simple MC. We then review the theory behind the Lasso regression method, and show how it can be used in the LMC algorithm. Finally, in section \ref{sec:benchmarks} LMC is benchmarked on a variety of examples. Proofs for all the theorems and lemmas are provided in appendix \ref{app:proofs}.

\subsection{Notation \& Assumptions}
Throughout the paper, bold letters represent vectors $\mvec x = (x_1,x_2,...,x_d)$, with $d$ the dimension. A lower case letter $x$ is a realisation of a random variable $X$, and in the case where $X$ is a multivariate random variable of dimension $d$, a realisation of it will be a vector $(x_1,x_2,...,x_d)$. For a random variable $X$ with probability density function (PDF) $\phi(x)$, we calculate the expectation value or mean with $\E[X] = \int_{\mathbb{R}}x\,\phi(x)\,dx\,$. We also use $\E[f(X)]$ or $\E[f]$ for the mean of the function $f$, whose input follows the distribution of $X$. The variance is defined as $\Var[X]=\E\Big[\big(X - \E[X]\big)^2\Big]$, the covariance $\Cov[X,Y] = \E\Big[\big(X - \E[X]\big)\big(Y - \E[Y]\big)\Big]$, the fourth central moment $m_4[X]=\E\Big[\big(X - \E[X]\big)^4\Big]$, and finally the multivariate second moment $m_{2,2}[X,Y]=\E\Big[\big(X - \E[X]\big)^2\big(Y - \E[Y]\big)^2\Big]$.

The function space $L^p\big(\mathbb{R}^d,\phi(\mvec x)d\mvec{x}\big)$ for $1\leq p<\infty$ is defined as the space of functions satisfying

$\left(\int_{\mathbb{R}^d}|f(\mvec x)|^p\phi(\mvec{x})d\mvec{x}\right)^{1/p}<\infty\,$.

The function $f$ is the ground truth, i.e. the expensive model, while \red{$\surr{n}$ is a cheap-to-evaluate surrogate model that was fitted to a training set of $n$ samples of $f$}. Similarly, $\mu_N$ and \red{$\mu_N\ntr{n}$} are the sample estimators for the mean of a sample set of size $N$, calculated with the true and surrogate model respectively. Also $\s2_N$ and \red{$\s2_N\ntr{n}$} are the sample estimators of the variance of a sample set of size $N$, computed with the true and surrogate models respectively. We write the two-level estimators as \red{$\mu_{N,M}\ntr{n}$ and $\s2_{N,M}\ntr{n}$}, and the LMC estimators as $\mathcal{M}_{N,M}$ and $\Sigma^2_{N,M}$. \red{It is assumed that the computational cost of training and evaluating $\surr{n}$ is negligible compared to the cost of evaluating $f$. Therefore the cost of an estimator is given by $N$, the number of evaluations of $f$.}

A normal distribution with mean $\mvec\mu$ and covariance matrix $\Sigma$ is written as $\mathcal{N}(\mvec\mu,\Sigma)$, and $U[a,b]$ is a uniform distribution between $a$ and $b$ with $a<b$.
With mean squared error (MSE) is defined as \red{$\MSE\left(\mu_N,\E[f]\right) = \E\left[\left(\mu_N-\E[f]\right)^2\right]$.}

Without loss of generality, we assume that all the sets of data used have been centred around zero, such that $\frac{1}{N}\sum_{i=1}^{N}f(\mvec x_i) = 0\,,\quad\text{and}\quad \frac{1}{N}\sum_{i=1}^{N} x_{ik} = 0\,,\quad k = 1,2,...,d\,$.

\red{The following assumption is made for any surrogate model $\surr{n}$, and section \ref{sec:lasso_in_estimators} discusses its validity.}

\begin{assumption}
    \red{Let $\surr{n}$ be a surrogate model that has been fitted on a training set of $n$ inputs $\mvec{x}_1,\mvec{x}_2,...,\mvec{x}_n$ sampled from i.i.d. random variables distributed as $X$, and $n$ outputs $f(\mvec{x}_1),f(\mvec{x}_2),...,f(\mvec{x}_n)$. Then the following inequalities are satisfied}

\begin{subequations}
    \label{eq:assumptions}
    \begin{align}
        &\Var\left[f-\surr{n}\right] \leq ~\Var\left[f\right]\,, \label{eq:assumption_a}\\
        \begin{split}
        & m_{2,2}\left[f+\surr{n}, f-\surr{n}\right] + \frac{1}{N-1}\Var\left[f+\surr{n}\right]\Var\left[f-\surr{n}\right]\\
        & \quad- \frac{N-2}{N-1}\left(\Var\left[f\right]-\Var\left[\surr{n}\right]\right)^2 \leq ~ m_4[f] - \frac{N-3}{N-1}{\Var}^2[f]\,,\label{eq:assumption_b}
        \end{split}
    \end{align}
\end{subequations}

for any integers $n>0$ and $N>3$.
\end{assumption}

\section{Theory}
\subsection{Two-Level Monte Carlo}
\label{sec:two-level_estimators}
The proposed method, LMC, is based on \red{multilevel and multifidelity Monte Carlo methods \cite{giles_multilevel_2008,Peherstorfer2016,peherstorfer_survey_2018}. As with any MC method,} one wants to estimate $\E[f]$ for some function $f$. Several models $f_1,f_2,...,f_L$ are available that approximate $f$, which have increasing cost and increasing accuracy, i.e. $f_L$ is the most accurate and expensive model, while $f_1$ is computationally cheap but inaccurate. \red{The difference in naming between \textit{multilevel} and \textit{multifidelity} methods comes from how these levels of accuracy are defined: in multilevel Monte Carlo the levels are obtained by coarsening or refining a grid, or changing the step-size of the integrator, whereas in multifidelity Monte Carlo the different functions are given by any general lower-order model such as data-fit surrogates or projection-based models. In both cases, however, }the goal of the method is to reduce the overall computational cost of computing $\E[f]$ with respect to traditional MC, by optimally balancing the amount of evaluations at each level of accuracy.

Originally MLMC was developed for estimating only the mean of a distribution, but in more recent years it has been extended \cite{bierig_estimation_2016, krumscheid_quantifying_2020, qian_multifidelity_2018} to estimate higher order moments of the distribution, which are necessary for UQ.

In the following paragraphs, a two-level version of M\red{F}MC is derived, in which \red{two levels of accuracy are considered}: the expensive, unbiased, true model $f$, and a computationally cheap, biased, surrogate model \red{$\surr{n}$}. 

\subsubsection{Mean Estimator}
\label{sec:mean_estimator}

Let $f(\mvec x)\in L^2\big(\mathbb{R}^d,\phi(\mvec x)d\mvec{x}\big)$ be a function, whose input is distributed according to the multivariate $d$-dimensional random variable $X$, with probability density function (PDF) $\phi(\mvec x)$. \red{Let $\mvec x_1,\mvec x_2,...,\mvec x_N$ and $\mvec z_1,\mvec z_2,...,\mvec z_M$ be two sets of input samples of size $N$ and $M$, drawn from the i.i.d. random variables distributed as $X$.} The aim is to estimate the mean $\E[f]$, with the minimum number of evaluations of the function $f$. The simple MC estimator is

\begin{equation}
    \mu_N = \frac{1}{N}\sum_{i=1}^N f(\mvec x_i)\,.
    \label{eq:MC_mean}
\end{equation}

The mean squared error of the estimator is 

\begin{equation}
    \MSE\Big(\mu_N\red{,}\E[f]\Big) = \frac{\Var[f]}{N}\,.
    \label{eq:MC_mean_MSE}  
\end{equation}

Therefore it is an unbiased estimator since $\lim_{N\rightarrow\infty}\mu_N = \E[f]$. With this method we require $N = \frac{\Var[f]}{\MSE}$ samples to obtain an estimation with a mean squared error of $\MSE$.

Now let \red{$\surr{N}\in L^2\big(\mathbb{R}^d,\phi(\mvec x)d\mvec{x}\big)$} be a surrogate model that \red{was trained on $N$ evaluations of $f$}, and is much cheaper to evaluate. Using this surrogate model to compute the sample mean, the estimator is

\begin{equation}
    \red{\mu_M\ntr{N} = \frac{1}{M}\sum_{i=1}^M \surr{N}(\mvec z_i)\,.}
    \label{eq:surrogate_mean}
\end{equation}

\red{Note that this estimator has the same cost as \eqref{eq:MC_mean} since the number of evaluations of $f$ is the same.} The error is

\begin{equation}
    \red{\MSE\Big(\mu\ntr{N}_{M},\E[f]\Big) = \left(\E[\surr{N}] - \E[f]\right)^2 + \frac{\Var\left[\surr{N}\right]}{M}\,.}
\end{equation}

The first term is the bias, and the second term is the variance. The variance term quickly vanishes if we assume that $\surr{N}$ has a negligible runtime, and that thus $M\rightarrow\infty$. However, the bias term can only be reduced by \red{increasing the training set size $N$} (and hence improving the accuracy of the surrogate model), which might be impossible or computationally demanding. \red{This is especially problematic in high-dimensional cases, since the volume of the input space to be sampled increases exponentially with $d$ \cite{koppen_curse_2000}, and unless the training set was large $N\gg d$ the surrogate will be heavily biased. Additionally, even if the training set were large and $N\rightarrow\infty$, the bias would in general not decay to zero due to model bias. }

Let us now combine the surrogate model with $f$ into a two-level estimator. \red{For this, assume that the set of $N$ samples is split into a \textit{training} subset $\mvec x_1,\mvec x_2, ...,\mvec x_n$ of size $n$ with $1\leq n\leq N$, and an \textit{evaluation} subset $\mvec x_{n+1},\mvec x_{n+2}, ...,\mvec x_N$ of size $N-n$.} The training samples are used to fit a surrogate $\surr{n}$. Then the two-level estimator reads

\begin{equation}
    \red{\mu_{N-n,M}\ntr{n} = \frac{1}{M}\sum_{i=1}^{M} \surr{n}(\mvec z_i) + \frac{1}{N-n}\sum_{i=n+1}^N f(\mvec x_i) - \surr{n}(\mvec x_i) = \mu\ntr{n}_M + \mu_{N-n} - \mu_{N-n}\ntr{n}\,.}
    \label{eq:LMC_mean}
\end{equation}

\red{As with the previous estimator, the cost is $N$, but in this case} it is an unbiased estimate since the error

\begin{equation}
    \red{\MSE\Big(\mu_{N-n, M}\ntr{n}, \E[f]\Big) = \frac{\Var\left[\surr{n}\right]}{M} +
    \frac{\Var\left[f - \surr{n}\right]}{N-n}}
    \label{eq:LMC_mean_MSE}
\end{equation}

has variance terms, but no bias term. Here again we assume that $\surr{n}$ has negligible runtime and that $M\rightarrow\infty$, and thus the first term vanishes. \red{Then, the following statement can be made regarding the MSE of simple MC and of the two-level estimator:}

\begin{equation}
   \red{\lim_{M\rightarrow\infty}\MSE\Big(\mu_{N-n, M}\ntr{n}, \E[f]\Big) \leq \MSE\Big(\mu_N\red{,}\E[f]\Big)\quad\iff\quad\frac{N}{N-n}\leq\frac{\Var[f]}{\Var\left[f - \surr{n}\right]}\,.}
    \label{eq:mean_condition}
\end{equation} 

Therefore, for a given computational budget $N$, the two-level estimator is equally or more accurate than simple MC if and only if $n$ and $\surr{n}$ are such that \eqref{eq:mean_condition} is satisfied. Note that the fraction on the far right of \eqref{eq:mean_condition} is guaranteed to be larger than $1$ by assumption \ref{eq:assumption_a}. Also in this fraction note that the denominator could in principle be zero, however we assume that this will never happen due to the model bias of $\surr{n}$.


\subsubsection{Variance Estimation}

Let $f(\mvec x)\in L^4\big(\mathbb{R}^d,\phi(\mvec x)d\mvec{x}\big)$. Let there be two sets of input samples of size $N$ and $M$, as in section \ref{sec:mean_estimator}. Then the simple MC estimator for the variance is 

\begin{equation}
    \sigma^2_N = \frac{1}{N-1}\sum_{i=1}^N \left(f(\mvec x_i) - \sum_{j=1}^N \frac{f(\mvec x_j)}{N}\right)^2\,,
    \label{eq:MC_var}
\end{equation}

which is unbiased and has an error

\begin{equation}
    \MSE\Big(\sigma^2_N, \Var[f]\Big) = \frac{1}{N}\left(m_4[f] - \frac{N-3}{N-1}{\Var}^2[f]\right)\,.
    \label{eq:MC_var_MSE}
\end{equation}

Using a surrogate model \red{$\surr{N}\in L^4\big(\mathbb{R}^d,\phi(\mvec x)d\mvec{x}\big)$} trained on $N$ evaluations of $f$, the variance estimator is

\begin{equation}
   \red{{\s2}_M\ntr{N} = \frac{1}{M-1}\sum_{i=1}^M \left(\surr{N}(\mvec x_i) - \sum_{j=1}^M \frac{\surr{N}(\mvec x_j)}{M}\right)^2\,,}
    \label{eq:surrogate_var}
\end{equation}

which has an error

\begin{equation*}
    \MSE\Big({\s2}_M\ntr{N}, \Var[f]\Big) = \Big(\Var[\surr{N}] - \Var[f]\Big)^2+ \frac{1}{M}\left(m_4[\surr{N}] - \frac{M-3}{M-1}{\Var}^2[\surr{N}]\right)\,.
\end{equation*}

The surrogate estimation of the variance is biased. The second term vanishes if $M\rightarrow\infty$, \red{while the first term is affected by model bias, and decays slowly due to the curse of dimensionality.}

\red{Now assume that the $N$ input samples are split into a subset of $n$ \textit{training} samples which are used to fit a surrogate $\surr{n}$, and a subset of $N-n$ \textit{evaluation} samples.} Then the two-level estimator for the variance is

\begin{equation}
       \red{ {\s2}\ntr{n}_{N-n,M} = {\s2}_M\ntr{n} + \sigma^{2}_{N-n} - {\s2}_{N-n}\ntr{n}\,,}
    \label{eq:LMC_var}
\end{equation}

\red{This estimator has a computational cost of $N$ and is unbiased.} The error is

\begin{equation}
\red{
    \begin{split}
        \MSE&\left({\s2}\ntr{n}_{N-n,M}, \Var[f]\right) \\
        &=\Var\left[{\s2}\ntr{n}_{M}\right] + \Var\left[\sigma^2_{N-n} - {\s2}\ntr{n}_{N-n}\right]\\
        &=\frac{1}{M}\left(m_4[\surr{n}] - \frac{M-3}{M-1}{\Var}^2[\surr{n}]\right) \\
        &\quad + \frac{1}{N-n}\Big(m_{2,2}\left[f+\surr{n}, f-\surr{n}\right] + \frac{1}{N-n-1}\Var[f+\surr{n}]\Var[f-\surr{n}] \\
        &\hspace{2cm} - \frac{N-n-2}{N-n-1}\left(\Var[f]-\Var[\surr{n}]\right)^2\Big)\,.
    \end{split}
    }
    \label{eq:LMC_var_MSE}
\end{equation}

Similarly to the mean estimator, the error \eqref{eq:LMC_var_MSE} has two variance terms. We assume that the first term vanishes as \red{$M\rightarrow\infty$}. \red{Then, the following statement can be made regarding the MSE of simple MC and of the two-level estimator:}

\begin{equation}
\red{
    \begin{split}
    & \lim_{M\rightarrow\infty}\MSE\left({\s2}\ntr{n}_{N-n,M}, \Var[f]\right) \leq \MSE\Big(\sigma^2_N, \Var[f]\Big)\quad\iff\\
    &\frac{N}{N-n}\leq \Big(m_4[f] - \frac{N-3}{N-1}{\Var}^2[f]\Big) \Big(m_{2,2}\left[f+\surr{n}, f-\surr{n}\right] \\
    & \hspace{1.2cm}+ \frac{1}{N-n-1}\Var[f+\surr{n}]\Var[f-\surr{n}] - \frac{N-n-2}{N-n-1}\left(\Var[f]-\Var[\surr{n}]\right)^2\Big)^{-1}\,.
    \end{split}
    }
    \label{eq:var_condition}
\end{equation}

Therefore, for a given $N$, the two-level estimator is equally or more accurate than simple MC if and only if $n$ and $\surr{n}$ are such that \eqref{eq:var_condition} is satisfied.

\subsection{Lasso Monte Carlo (LMC)}
\label{sec:LMC}

\red{Consider a fixed computational budget of $N$ evaluations of $f$, and a fixed choice of surrogate model. Then, the mean squared errors of the two-level estimators (\ref{eq:LMC_mean_MSE}, \ref{eq:LMC_var_MSE}) are controlled by $n$. There is a trade-off in the choice of this number, a larger $n$ increases the training set size and potentially reduces the bias of $\surr{n}$, while it also increases the variance of the estimator since the number of evaluation samples $N-n$ is decreased. In the default implementation of MFMC (henceforth referred to as \textit{static MFMC}), the choice of $n$ and the training of the surrogate are done beforehand, independently of the MSE. In recent years \textit{adaptive MFMC} \cite{peherstorfer_adaptive_2019} (sometimes also referred to as \textit{context-aware surrogate models} \cite{alsup_context-aware_2023}) were introduced. With this approach, an upper bound of the MSE is estimated, and used to calculate the optimal $n$ that minimises the errors. Both of these approaches converge faster than simple MC \cite{peherstorfer_survey_2018, peherstorfer_adaptive_2019}, however there is no guarantee that the MFMC estimates are more accurate than those of simple MC. Indeed, for a given $N$, it is possible that there exists no choice of $n$ in $\{1,...,N\}$ such that conditions (\ref{eq:mean_condition}, \ref{eq:var_condition}) are satisfied.}

\red{A possible way to guarantee that the two-level estimators are more accurate than simple MC, is by considering a larger computational budget of $n+N$. Let $\mvec x_1, \mvec x_2,...,\mvec x_n$ be the training set and $\mvec x_{n+1}, \mvec x_{n+2},...,\mvec x_{n+N}$ be the evaluation set. Then the estimators (\ref{eq:LMC_mean},\ref{eq:LMC_var}) become $\mu\ntr{n}_{N,M}$ and ${\s2}\ntr{n}_{N,M}$, which would have the following errors:}

\begin{equation}  
\red{
    \begin{split}
        \MSE&\Big(\mu_{N, M}\ntr{n}, \E[f]\Big) = \frac{\Var\left[f - \surr{n}\right]}{N} + \frac{\Var\left[\surr{n}\right]}{M}\\
        \MSE&\left({\s2}\ntr{n}_{N,M}, \Var[f]\right) = \frac{1}{N}\Big(m_{2,2}\left[f+\surr{n}, f-\surr{n}\right] \\
        &\hspace{2cm} + \frac{1}{N-1}\Var[f+\surr{n}]\Var[f-\surr{n}] \\
        &\hspace{2cm} - \frac{N-2}{N-1}\left(\Var[f]-\Var[\surr{n}]\right)^2\Big)\\
        & \hspace{2cm} + \frac{1}{M}\left(m_{4}[\surr{n}]-\frac{M-3}{M-1}{\Var}^2[\surr{n}]\right)\,.
    \end{split}
}
    \label{eq:LMC_best_MSE}
\end{equation}

\red{These estimators are guaranteed to be more accurate than simple MC due to assumptions \ref{eq:assumptions}:}

\begin{equation}
\red{
    \begin{split}
        &\lim_{M\rightarrow\infty}\MSE\Big(\mu_{N, M}\ntr{n}, \E[f]\Big) \leq \MSE\Big(\mu_N\red{,}\E[f]\Big)\quad\iff\quad \text{Assumption }\ref{eq:assumption_a}\,,\\
        &\lim_{M\rightarrow\infty}\MSE\left({\s2}\ntr{n}_{N,M}, \Var[f]\right) \leq \MSE\Big(\sigma^2_N, \Var[f]\Big)\quad\iff\quad\text{Assumption }\ref{eq:assumption_b}\,.
    \end{split}
}
    \label{eq:conditions}
\end{equation}

\red{This approach, despite guaranteeing higher accuracy than simple MC, has a larger computational cost as it requires $N+n$ evaluations of $f$. One could naively  reduce the costs by reusing the same $N$ samples for training $\surr{N}$ and evaluating the estimators, thus removing the need for the additional $n$ training samples. This, however, would introduce bias in the estimator since the evaluation and training samples must come from i.i.d. random variables. The LMC algorithm introduced in this work presents an alternative method, with computational cost $N$, and MSE similar to \eqref{eq:LMC_best_MSE}. To avoid introducing bias,} LMC splits the training set into $S$ subsets, and trains $S$ surrogate models. The full algorithm is detailed in (Algorithm \ref{alg:LMC}). Note that the LMC algorithm has \textit{Lasso} in its name, because in this paper the surrogate model $\surr{n}$ used in the algorithm was exclusively a Lasso model. However, algorithm \ref{alg:LMC} does not make any reference to Lasso and is in principle agnostic to the method used to train $\surr{n}$.


\begin{algorithm}[H]
\caption{The Lasso Monte Carlo Algorithm}\label{alg:LMC}
\begin{algorithmic}[1]
\Require \red{Input sets $V=\{\mvec x_1,\mvec x_{2}, ..., \mvec x_N\}$ and $W=\{\mvec z_{1},\mvec z_{2}, ..., \mvec z_{M}\}$, and an integer $S$ that divides $N$.}
\Ensure $N \ll M$

\State Compute the outputs of set $V$: $f(\mvec x_1),f(\mvec x_2), ..., f(\mvec x_N)$.
\State Compute $\mu_{N},\sigma^2_{N}$ with the simple MC estimators (\ref{eq:MC_mean}, \ref{eq:MC_var}).
\label{step:splits}

\For{$\smalls=1 \ldots S$}
    \State \red{Define an \textit{evaluation} subset of size $\frac{N}{S}$ as $V_{eval}=\{\mvec x_{N(\smalls-1)/S+1},\mvec x_{N(\smalls-1)/S+2}, ..., \mvec x_{N\smalls/S}\}\subset V$.}
    \State \red{The remaining samples define the \textit{training} subset $V_{train}= V\setminus V_{eval}$ of size $n= N\frac{S-1}{S}$.}
    \State \red{Fit $\surr{n}_{\smalls}$ to $V_{train}$.}
    \State \red{With $\surr{n}_{\smalls}$, compute the outputs of $W$ and $V_{eval}$.}
    \State \red{Use $W$ and $V_{eval}$ to }compute $\left(\mu\ntr{n}_{N/S,M}\right)_{\smalls}$ and $\left({\s2}\ntr{n}_{N/S,M}\right)_{\smalls}$ with the two-level estimators (\ref{eq:LMC_mean},\ref{eq:LMC_var}).
\EndFor
\State Compute the LMC mean and variance, by averaging out the two-level estimations:

\begin{equation}
    \mathcal{M}_{N,M} = \frac{1}{S}\sum_{\smalls=1}^S\left(\mu\ntr{n}_{N/S,M}\right)_{\smalls}\,,\quad\text{and}\quad \Sigma^2_{N,M} = \frac{1}{S}\sum_{\smalls=1}^S\left({\s2}\ntr{n}_{N/S,M}\right)_{\smalls}\,.
    \label{eq:final_LMC}
\end{equation}
\end{algorithmic}
\end{algorithm}

This algorithm only requires $N$ evaluations of $f$, and therefore has the same cost as simple MC. Note that the LMC estimators \eqref{eq:final_LMC} differ from the two-level estimators (\ref{eq:LMC_mean}, \ref{eq:LMC_var}), since LMC computes the average of several smaller two-level estimators. It can be shown that, under certain assumptions, the LMC estimators have similar accuracy to \eqref{eq:LMC_best_MSE}, in which case they are guaranteed to be equally or more accurate than simple MC (see eq. \ref{eq:conditions}). Furthermore, in the benchmarks section (sec. \ref{sec:benchmarks}) LMC is found to be more accurate than simple MC, static MFMC, and adaptive MFMC for all $N$.

\begin{theorem}[LMC Accuracy] Let $\surr{n}_1=\surr{n}_2=...=\surr{n}_S$. Then the LMC estimators $\mathcal{M}_{N,M}$ and $\Sigma^2_{N,M}$ defined by \eqref{eq:final_LMC}, have the errors $\lim_{M\rightarrow\infty}\MSE\left(\mu\ntr{n}_{N,M}, \E[f]\right) = \lim_{M\rightarrow\infty}\MSE\left(\mathcal{M}_{N,M}, \E[f]\right)$ and $\lim_{M\rightarrow\infty}\MSE\left({\s2}\ntr{n}_{N,M}, \Var[f]\right) = \lim_{M\rightarrow\infty}\MSE\left(\Sigma^2_{N,M}, \Var[f]\right) + \mathcal{O}\left(N^{-2}\right)$.
\label{thm:LMC_equal}
\end{theorem}

\begin{remark}
\label{rem:equal_surrogates}
    The condition $\surr{n}_1=\surr{n}_2=...=\surr{n}_S$ occurs if a surrogate model $\surr{n}$ is independent of the elements of the training set, and only depends on the size $n$ of the training set. This occurs when the training set is large, or $\surr{n}$ is a model with small variance.
\end{remark}

\begin{figure}[H]
    \centering
    \includegraphics[width=\textwidth]{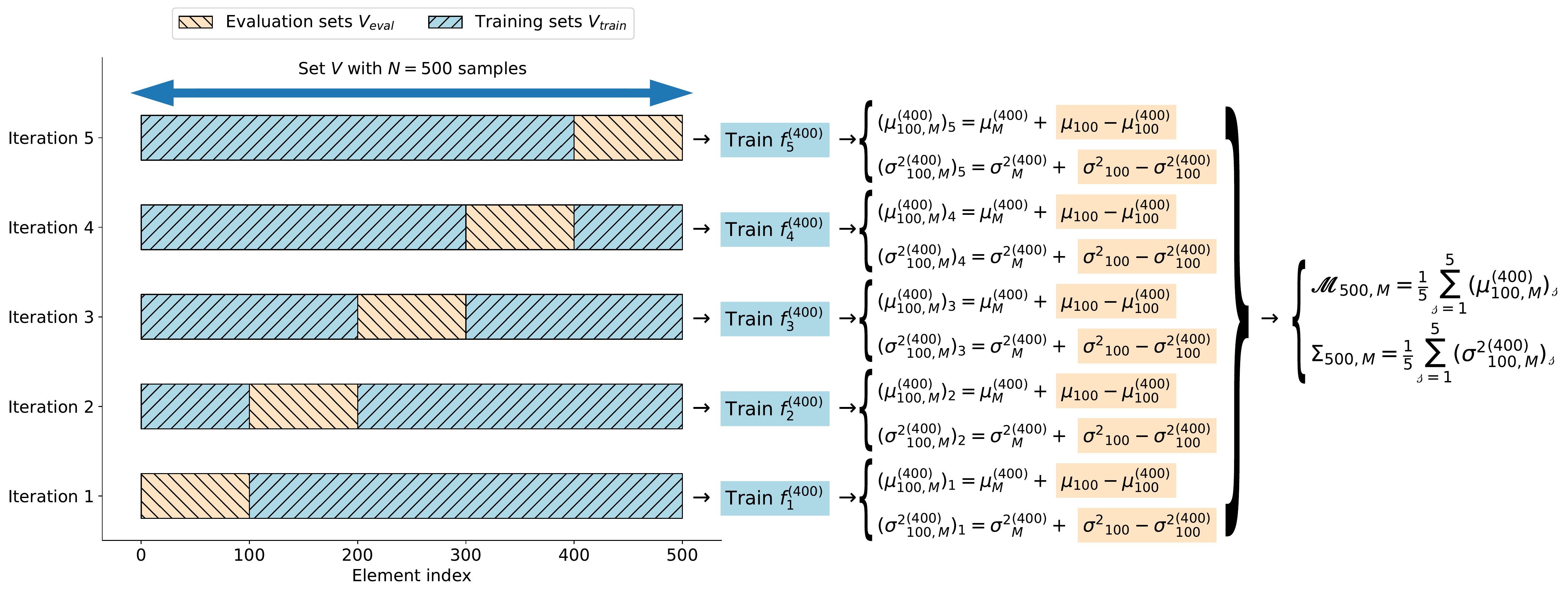}
    \caption{\red{Illustration of the LMC algorithm \ref{alg:LMC}, with $S=5$ and $N=500$. Accordingly, the evaluation sets $V_{eval}$ have size $100$, and the training sets $V_{train}$ have size $n=400$.}}
    \label{fig:diagram_CV}
\end{figure}

\red{As a more intuitive interpretation of LMC, we point out that the definition of $V_{eval}$ and $V_{train}$ in the for-loop of algorithm \ref{alg:LMC}, is equivalent to performing an S-fold split on the set $V$, as described in figure \ref{fig:diagram_CV}.}

\subsection{Lasso Regression}
\label{sec:Lasso}

\red{In remark \ref{rem:equal_surrogates} it was pointed out that a surrogate model with a small variance is required for the LMC method. For this reason, in this work it was decided to use a Lasso regression model \cite{tibshirani_regression_1996}, as it has a regularisation parameter $\lambda$ for straight-forward tuning of the model variance. This section introduces Lasso.}

The Lasso regression method \cite{tibshirani_regression_1996} estimates the weights $\mvec{\beta}\in\mathbb{R}^d$ of a linear model of the type
\begin{equation}
    \red{\surr{n}}(\mvec x) = \mvec\beta\cdot\mvec x,
    \label{eq:lasso}
\end{equation}

such that $\surr{n}$ approximates $f$. This method estimates $\mvec\beta$ by minimising a loss function on a training set $\mvec x_1, f(\mvec x_1)$, $\mvec x_2, f(\mvec x_2)$, ..., $\mvec x_{n}, f(\mvec x_{n})$ of size $n$. The Lasso loss function is

\begin{equation}
    \mathcal{L}(\mvec\beta) = \frac{1}{2}\sum_{i=1}^{n}\left(f(\mvec x_i) - \mvec\beta\cdot\mvec x_i \right)^2 + \lambda||\mvec{\beta}||_1\,,
    \label{eq:lasso_loss}
\end{equation}

where $\lambda>0$ is a chosen \textit{regularisation} constant, and $||\mvec\beta||_1$ is the $L^1$-norm

\begin{equation*}
    ||\mvec\beta||_1 = \sum_{k=1}^d|\beta_k|\,.
\end{equation*}

The first term of the Lasso loss function \eqref{eq:lasso_loss} is the usual ordinary least squares (OLS) loss function. In an overdetermined system, i.e. with $n>d$, the OLS loss function is strictly convex and has a global minimum at

\begin{equation}
    \mvec\beta^{\text{OLS}} = \red{\left(A^TA\right)^{-1}A^T\mvec{f}\,,}
    \label{eq:OLS_solution}
\end{equation}

where $A\in\mathbb{R}^{n\times d}$ is a nonsingular matrix containing the input training data, and $\mvec{f} = \left(f(\mvec x_1), f(\mvec x_2), ..., f(\mvec x_{n})\right)$ is the output training data, or in general the quantities of interest.

In an underdetermined system, i.e. $n<d$, the OLS loss is weakly convex, and therefore the problem is ill-posed and has multiple minima. Moreover, if $A$ is close to collinear the OLS solution \eqref{eq:OLS_solution} has large variance. The Lasso approach simultaneously solves the problems of underdetermination and large variance by adding a second term to the loss \eqref{eq:lasso_loss}: the regularisation term, also referred to as the constraint or shrinkage term. This term adds a penalty on the complexity of the model. More specifically, the Lasso regularisation has the effect of shrinking the model weights $\mvec\beta$ towards zero, and forcing some of them to be exactly zero, thus yielding a sparse model. In this sense Lasso can be seen as a variable selection strategy \cite{tibshirani_regression_1996}, in which the least important inputs are suppressed, and the dimensionality of the problem is reduced. The regularisation parameter $\lambda$, chosen by the user, controls the amount of shrinkage: for large $\lambda$ the weight vector $\mvec\beta$ is small in magnitude and highly sparse, and if $\lambda$ is large enough the weight vector only contains zeros (see lemma \ref{lem:lambda_max}). Oppositely, if $\lambda$ is small the weight vector approaches the OLS solution, i.e. $\lambda\rightarrow 0^+$ leads to $\mvec\beta\rightarrow\mvec\beta^{\text{OLS}}$. 

\begin{lemma}[Lasso with Null Weights]
\label{lem:lambda_max}
Let $\{x_{ik}\}_{i=1,2,...,n}^{k=1,2,...,d}$ be the input data, drawn from some continuous distribution, and $f(\mvec x_1), f(\mvec x_2),...,f(\mvec x_{n})$ the outputs. Let $\mvec\beta$ be the weight vector that minimises the Lasso loss function \eqref{eq:lasso_loss}, with a regularisation parameter $\lambda$. Then, 

\begin{equation}
    \mvec\beta = (0,0,...,0)\quad\iff\quad\lambda\geq\lambda_{max} \quad\text{with}\quad \lambda_{max} = \max_{k=1,2,...,d}\left|\sum_{i=1}^{n}x_{ik}f(\mvec{x_i})\right|\,.
    \label{eq:lambda_max}
\end{equation}
\end{lemma}

\begin{remark}
The minimum of the Lasso loss \eqref{eq:lasso_loss} does not have a closed-form expression, and therefore an optimisation algorithm is required for finding a solution. In this work we use the scikit-learn \cite{pedregosa_scikit-learn_2011} implementation of Lasso, which uses the coordinate descent algorithm \cite{fu_penalized_1998,wu_coordinate_2008}.
\end{remark}

\subsection{Lasso in \red{LMC} }
\label{sec:lasso_in_estimators}

\red{As per theorem \ref{thm:LMC_equal} and conditions \eqref{eq:conditions}, the LMC estimators \eqref{eq:final_LMC} are equally or more accurate than simple MC if the surrogate model $\surr{n}$ satisfies assumptions \eqref{eq:assumptions}. Up until now these assumptions have been taken for granted, and in this section we discuss their validity when the surrogate model is in fact a Lasso model, as described in section \ref{sec:Lasso}.}

\subsubsection{Lasso in the LMC Mean Estimator}

\red{The following theorem shows that a Lasso surrogate model satisfies assumption \eqref{eq:assumption_a} under mild conditions, and therefore the LMC estimator for the mean is almost always guaranteed to be equally or more accurate than simple MC.}

\begin{theorem}[Lasso in \red{Assumption \eqref{eq:assumption_a}}]
Let $f(\mvec x)\in L^2\left(\mathbb{R}^d,\phi(\mvec x)d\mvec{x}\right)$ be a function, whose input is a multivariate random variable $X = (X_1,X_2,...,X_d)$, with a PDF $\phi(\mvec{x})$ and nonsingular covariance matrix $\Sigma$. Let $\mvec x_1, f(\mvec{x}_1), \mvec x_2, f(\mvec{x}_2), ..., \mvec{x_n}, f(\mvec{x}_{n})$ be a random sample of input-outputs, called the training set. Let $\surr{n}$ be a surrogate model that was trained on this training set by minimising the Lasso loss function \eqref{eq:lasso_loss}, with a regularisation parameter $\lambda$. Then

\begin{enumerate}[label=\emph{\alph*})]
    \item in the asymptotic limit $n\rightarrow\infty$, $f$ and $\surr{n}$ satisfy assumption \eqref{eq:assumption_a} for any regularisation parameter $|\lambda|<\infty$,
    \item in the non-asymptotic limit $n<\infty$, there exists a regularisation parameter $0<\lambda<\infty$ such that $f$ and $\surr{n}$ satisfy assumption \eqref{eq:assumption_a}
\end{enumerate}
\label{thm:Lasso_Mean}
\end{theorem}




\begin{remark}
Despite the existence of a $\lambda$ that satisfies \eqref{eq:assumption_a} in the non-asymptotic limit, estimating such a $\lambda$ is a nontrivial problem when $n<\infty$. In general, estimating $\lambda$ with theoretical guarantees requires prior knowledge of $f$. In this work we use 5-fold cross validation (CV) \cite{efron_introduction_1993} (except in benchmark \ref{sec:linear_benchmark}, see explanation therein), as suggested in the original Lasso paper \cite{tibshirani_regression_1996}, despite CV not having any theoretical guarantees in the non-asymptotic case. We note that there exist approaches that, under appropriate conditions of $\{x_{ik}\}$ and $f$, can compute the Lasso estimator with certain guarantees \cite{belloni_square-root_2011, chichignoud_practical_2016}.
\end{remark}

\subsubsection{Lasso in the LMC Variance estimator}

\red{If assumption \eqref{eq:assumption_b} is satisfied then the LMC estimator for the variance is guaranteed to be equally or more accurate than simple MC. However, in lemma \ref{lem:Lasso_fail} we show that, a Lasso model sometimes fails to satisfy this assumption.}

Nevertheless, as can be seen in section \ref{sec:benchmarks}, LMC is \red{found to be more accurate than simple MC and MFMC} in all presented cases. The success of LMC can be explained with the following heuristic argument: in UQ problems the input of $f(\mvec x)$ is the perturbation of some nominal value, i.e. $\mvec x = \mvec x_0 + \mvec\xi$, with $\mvec\xi$ a random value that represents the uncertainty of the input. Then, in the vicinity of $\mvec x_0$, the problem can be Taylor expanded into a linear function with some additional smaller terms $f(\mvec x) = f(\mvec x_0) + \mvec\xi\cdot\mvec\nabla f(\mvec x_0) + \mathcal{O}\left(||\mvec\xi||^2\right)\,$. In theorem \ref{thm:Lasso_Variance} we show that if $f$ is a linear function with added random noise, then Lasso does satisfy assumption \ref{eq:assumption_b}, in which case LMC is guaranteed to \red{have an equally or smaller MSE} than simple MC.

\begin{lemma}[Lasso in \red{Assumption \eqref{eq:assumption_b}}]
Let $f(\mvec x)\in L^4\left(\mathbb{R}^d,\phi(\mvec x)d\mvec{x}\right)$ be a function, whose input is a multivariate random variable $X = (X_1,X_2,...,X_d)$. Let $\mvec x_1, f(\mvec{x}_1)$ $, \mvec x_2, f(\mvec{x}_2),$ $...,$ $\mvec{x_{n}}, f(\mvec{x}_{n})$ be a random sample of input-outputs, called the training set. Let $\surr{n}$ be a surrogate model that was trained on this training set by minimising the Lasso loss function \eqref{eq:lasso_loss}. Then $f$ and $\surr{n}$ in general do not satisfy assumption \eqref{eq:assumption_b}.
\label{lem:Lasso_fail}
\end{lemma}

\begin{theorem}[Lasso in \red{Assumption \eqref{eq:assumption_a}}, with a noisy linear function]
Let $f(\mvec x)\in L^4\left(\mathbb{R}^d,\phi(\mvec x)d\mvec{x}\right)$ be a function with the following properties:

\begin{enumerate}[label=\roman*)]
    \item The input vector contains $d$ uncorrelated random variables $\mvec X = (X_1,X_2,...,X_d)$, with a joint PDF $\phi(\mvec{x})$, and a covariance matrix $\Sigma = \mathbb{1}_d$
    \item $f$ is a linear function with uncorrelated noise, i.e.
    \begin{equation}
        f(\mvec{x}) = \mvec\alpha\cdot\mvec x + \varepsilon\,,
        \label{eq:linear_noisy_function}
    \end{equation}
    with $\mvec{\alpha}$ the true weights of the function, and $\varepsilon$ a realisation of the random variable $\Epsilon$. The random variable $\Epsilon$ has zero mean and finite variance, and is independent of $X$.
\end{enumerate}
    Let $\mvec x_1, f(\mvec{x}_1), \mvec x_2, f(\mvec{x}_2), ..., \mvec{x_{n}}, f(\mvec{x}_{n})$ be a random sample of input-outputs, called the training set. Let $\surr{n}$ be a surrogate model that was trained on the training set by minimising the Lasso loss function \eqref{eq:lasso_loss}, with a regularisation parameter $\lambda$. Then
    
    \begin{enumerate}[label=\emph{\alph*})]
    \item in the asymptotic limit $n\rightarrow\infty$, $f$ and $\surr{n}$ satisfy assumption \ref{eq:assumption_b} for any regularisation parameter with $|\lambda|<\infty$,
    \item in the non-asymptotic limit $n<\infty$, there exists a regularisation parameter $0<\lambda<\infty$ such that $f$ and $\surr{n}$ satisfy assumption \eqref{eq:assumption_b}.
\end{enumerate}
\label{thm:Lasso_Variance}
\end{theorem}


\section{Numerical Examples}
\label{sec:benchmarks}

In the following examples the LMC algorithm \eqref{alg:LMC} was used, with $S=5$. All the calculations were carried out on single CPU processors of type Intel Xeon Gold 6152 with 2.10 GHz and 16 GB of RAM.

The following plots compare LMC to other methods, by comparing the relative error of the estimators for a given \red{computational budget $N$ (recall that $N$ is the number of evaluations of $f$, and the remaining costs are assumed to be negligible)}. To compute the errors, each method is evaluated $20$ times (or more), where each time a different seed is used to randomly generate the datasets. The relative errors are calculated with

\begin{equation}
    \begin{split}
        \text{Relative error on mean}&=\frac{1}{20}\sum_{r=1}^{20}\left|\frac{\mu(N,r) - \E[f]}{\E[f]}\right|,\\
        \text{Relative error on standard deviation}&=\frac{1}{20}\sum_{r=1}^{20}\left|\frac{\sqrt{\sigma^2(N,r)} - \sqrt{\Var[f]}}{\sqrt{\Var[f]}}\right|\,,\\
    \end{split}
    \label{eq:relative_errors}
\end{equation}

where $\mu(N,r)$ and $\sigma^2(N,r)$ are the estimators of a given method, using $N$ samples and the $r$-th initial seed. The error bars in the plots are given by the standard deviation of the relative errors \eqref{eq:relative_errors}. In the cases where the analytical expressions for $\E[f]$ and $\Var[f]$ are not known, they are approximated with the sample estimators (\ref{eq:MC_mean},~\ref{eq:MC_var}) and a large sample set.

\subsection{Linear Function}
\label{sec:linear_benchmark}
Let $f$ be a linear function with a large input dimension $d=400$:
\begin{equation*}
    \begin{cases}
        & f(\mvec x) = \mvec{\alpha}\cdot\mvec x,\\
        & \text{with }\mvec\alpha = \left(1, \frac{1}{2}, \frac{1}{5}, \frac{1}{10}, \frac{1}{20}, \frac{1}{50}, \frac{1}{100}, \frac{1}{100}, ..., \frac{1}{100}\right),
    \end{cases}
\end{equation*}

with $\text{dim}(\mvec\alpha) = d$ and with a normally distributed input $X\sim\mathcal{N}(\mvec 0,\mathbb{1}_d)$. 

\red{To perform a thorough benchmark, LMC is compared to all other methods discussed in this work, with an increasing computational budget $N$. Simple MC is applied as in eqs. (\ref{eq:MC_mean}, \ref{eq:MC_var}), and the \textit{Lasso} approach consists of a simple surrogate model approach as in eqs. (\ref{eq:surrogate_mean}, \ref{eq:surrogate_var}). The LMC is applied as in algorithm \ref{alg:LMC} with $S=5$. The \textit{Static MFMC} method is implemented with the two-level estimators (\ref{eq:LMC_mean}, \ref{eq:LMC_var}) with $n=0.8\cdot N$, and \textit{Adaptive MFMC} is applied in the same way, but the optimal $n$ is chosen by testing several values $n\in[0.1\cdot N, 0.2\cdot N, ..., 0.9\cdot N]$ and choosing the one with the smallest estimated MSE. Finally, \textit{Biased MFMC} uses the full set of $N$ samples for fitting $\surr{N}$ and then again for evaluating the two-level estimators. In all of the methods $M=10^5$, this ensures that the leading error in the two-level estimators comes from the $\bigO(N^{-1})$ term. For this benchmark, the $\lambda$ of the Lasso models is chosen such that the number of non-zero elements of $\mvec\beta$ is $\lfloor0.95\cdot N\rfloor$, where $\lfloor\,\rfloor$ is the floor function. Each model was evaluated 30 times with different seeds for generating the random input points.}

\begin{figure}[H]
    \centering
    \includegraphics[width=\textwidth]{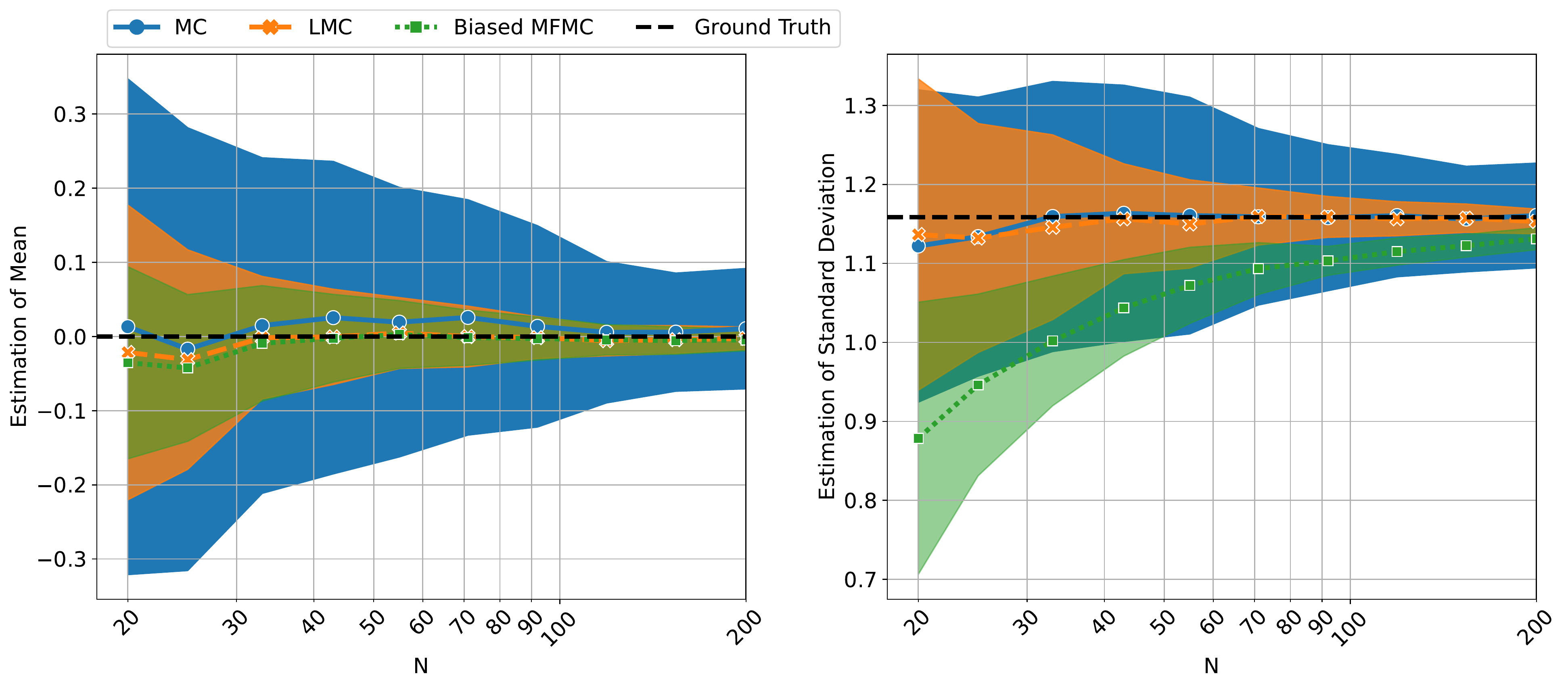}
    \caption{Estimation of the mean and standard deviation of $f(X)$, with respect to the \red{computational budget} $N$. The three models were run 30 times each with different initial seeds. The lines are the means of the 30 runs, and the coloured-in regions are bounded by the standard deviation of the 30 runs.}
    \label{fig:huge_linear_conv}
\end{figure}

Figure \ref{fig:huge_linear_conv} shows the comparison between simple MC, LMC, and Biased MFMC, being used to estimate the mean and standard deviation of $f(X)$. For the estimation of the mean, the LMC and Biased MFMC models perform similarly, indicating that the Lasso model has a small bias with respect to the mean. They both have a smaller MSE than simple MC. For the convergence of the standard deviation we see three different behaviours: 

\begin{itemize}
    \item the MC approach is unbiased and converges towards the true value, albeit \red{with a large MSE},
    \item the Biased MFMC approach is clearly biased. This is expected given that the training and evaluations sets are the same,
    \item the LMC approach is unbiased and \red{has a smaller MSE} than MC.
\end{itemize}

\begin{figure}[H]
    \centering
    \includegraphics[width=\textwidth]{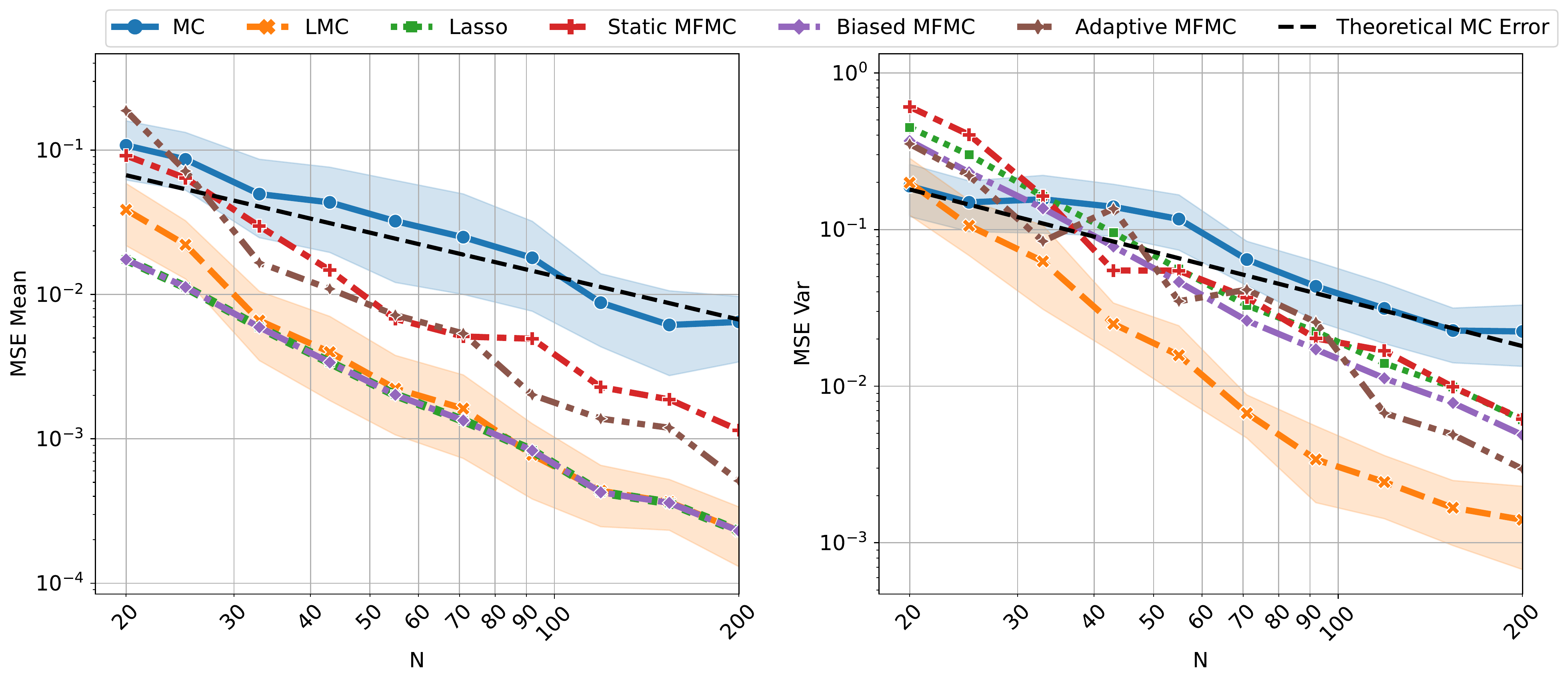}
    \caption{\red{Mean squared error on the estimation of the mean and variance} of $f(X)$, with respect \red{to the computational budget} $N$. Each line is the mean of \red{30} runs of a given estimator with randomly generated data. To improve readability, \red{the $95\%$ confidence intervals have been plotted only for the MC and LMC methods.}}
    \label{fig:huge_linear_errorconv}
\end{figure}

Figure \ref{fig:huge_linear_errorconv} shows the \red{mean squared error} of all methods. \red{It can be clearly seen that LMC is the most accurate method for the estimation of the mean and variance, and that it has an MSE smaller or equal than simple MC for all $N$, as one would expect from the theory (see thm. \ref{thm:LMC_equal}).}

\subsection{Sobol Function}
Let $f$ be the nonlinear function known as the Sobol function, with input dimension $d=400$:

\begin{equation}
    \begin{cases}
        & f(\mvec x) = \prod_{i=1}^d\frac{|4x_i-2| + c_i}{1+c_i},\\
        & \text{with }\mvec c = \left(1,2,5,10,20,50,100,500,500,...,500\right),
    \end{cases}
    \label{eq:sobol}
\end{equation}

where the input is sampled from $X_1,X_2,...,X_d$, i.i.d. uniform random variables, i.e. $X_k\sim U[0,1]\quad\forall k=1,2,...,d$. The Sobol function is a commonly used benchmark for UQ and sensitivity analysis (e.g. \cite{sudret_global_2008, konakli_global_2016, christos_data-driven_2019}), with unit mean, and variance analytically computed with $\sigma^2=\prod_{i=1}^d\Bigg(\frac{1}{3(1+c_i)^2}+1\Bigg)-1$.

\begin{figure}[H]
    \centering
    \includegraphics[width=\textwidth]{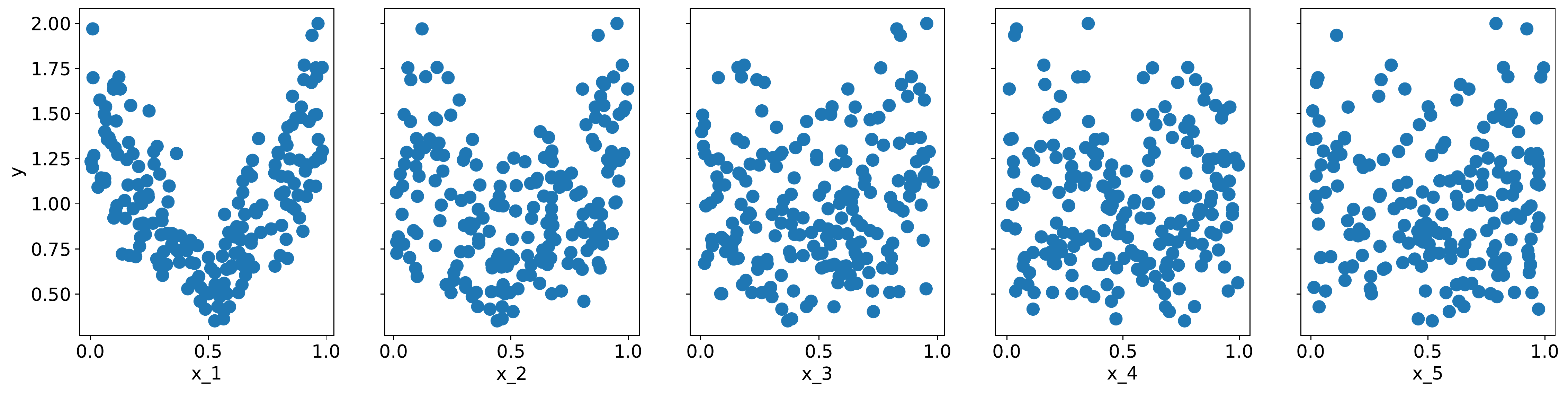}
    \caption{Output of the Sobol function with respect to the first 5 input dimensions, out of 400 dimensions in total. For these plots 200 sample points were used.}
    \label{fig:sobol_inputs}
\end{figure}

\red{Since} the Sobol function \eqref{eq:sobol} is symmetric around $\frac{1}{2}$ (see fig. \ref{fig:sobol_inputs}), \red{ fitting a Lasso (or other linear) model to it results in a constant model with $\mvec \beta = (0,0,...,0)$. In such a case the two-level estimators (\ref{eq:LMC_mean}, \ref{eq:LMC_var}) become exactly equal to the simple MC estimators, and no advantage would be gained from using LMC.}

Nevertheless this problem can be circumvented by adding a nonlinearity to the Lasso model. This is achieved with a nonlinear transformation $\zeta$ on the input, prior to fitting the surrogate model, which is equivalent to fitting the model $\surr{n} = \mvec\beta\cdot\mvec z$, with $\mvec z = \zeta(\mvec x)$. If $\zeta$ is adequately chosen, such that $\Cov[f,X_k]\neq 0$ for some $k$, the LMC estimator will no longer be equal to simple MC.

In practice, finding a good nonlinear transformation requires having some knowledge of $f$. In this case, the analytical expression of the Sobol function \eqref{eq:sobol} is known, thus one can intuitively choose the nonlinear transformation $\zeta(\mvec x) = |\mvec{x} - 0.5|$ where the subtraction and absolute value operations are applied element-wise to $\mvec x$. If $f$ is not known, one might come up with a similar transformation by plotting a few samples against the input dimensions, as in figure \ref{fig:sobol_inputs}.

A further remark regarding the Sobol function \eqref{eq:sobol} is that it is not a \textit{noisy linear function} of the type \eqref{eq:linear_noisy_function} discussed in thm. \ref{thm:Lasso_Variance}. Therefore, it is not guaranteed that the LMC variance estimation will \red{be more accurate} than simple MC, since $f$ and $\surr{n}$ might not satisfy \eqref{eq:assumption_b}, as explained in lemma \ref{lem:Lasso_fail}. In a real scenario, with unknown $f$, LMC and simple MC are simultaneously evaluated, the MSE is estimated, and the method with the smallest MSE is used.

\begin{figure}[H]
    \centering
    \includegraphics[width=\textwidth]{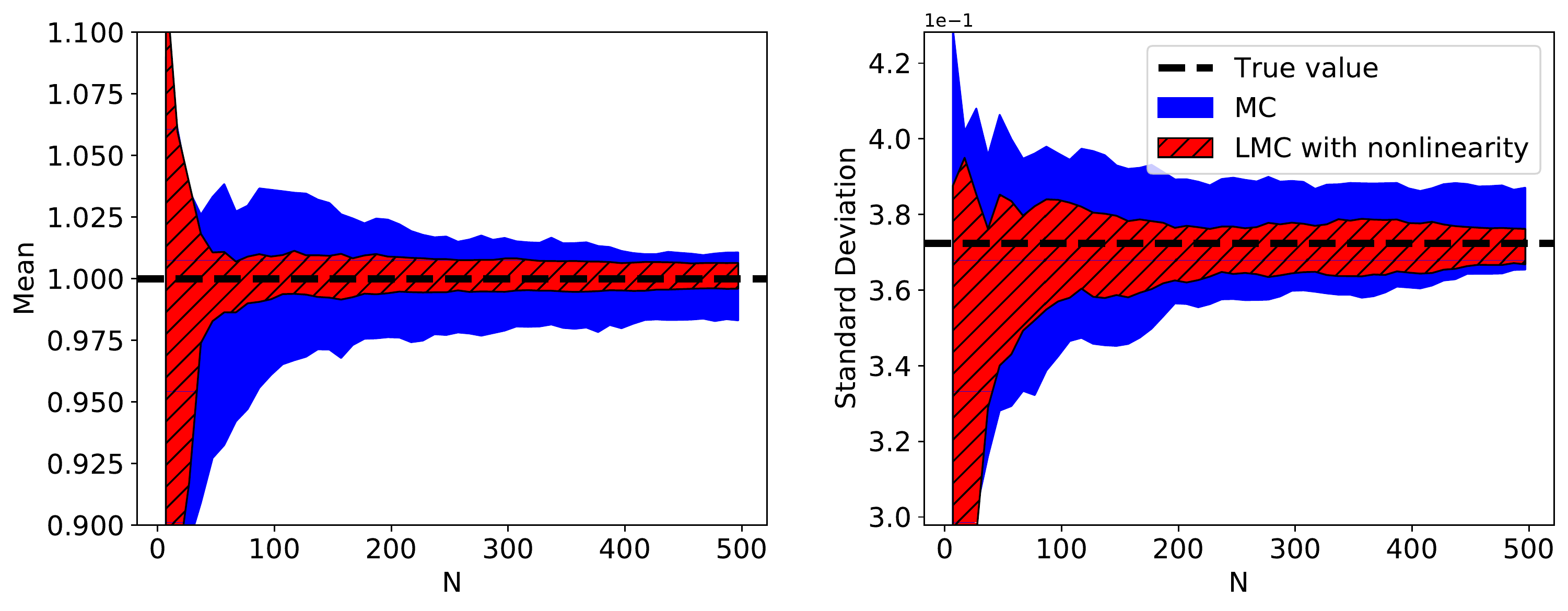}
    \includegraphics[width=\textwidth]{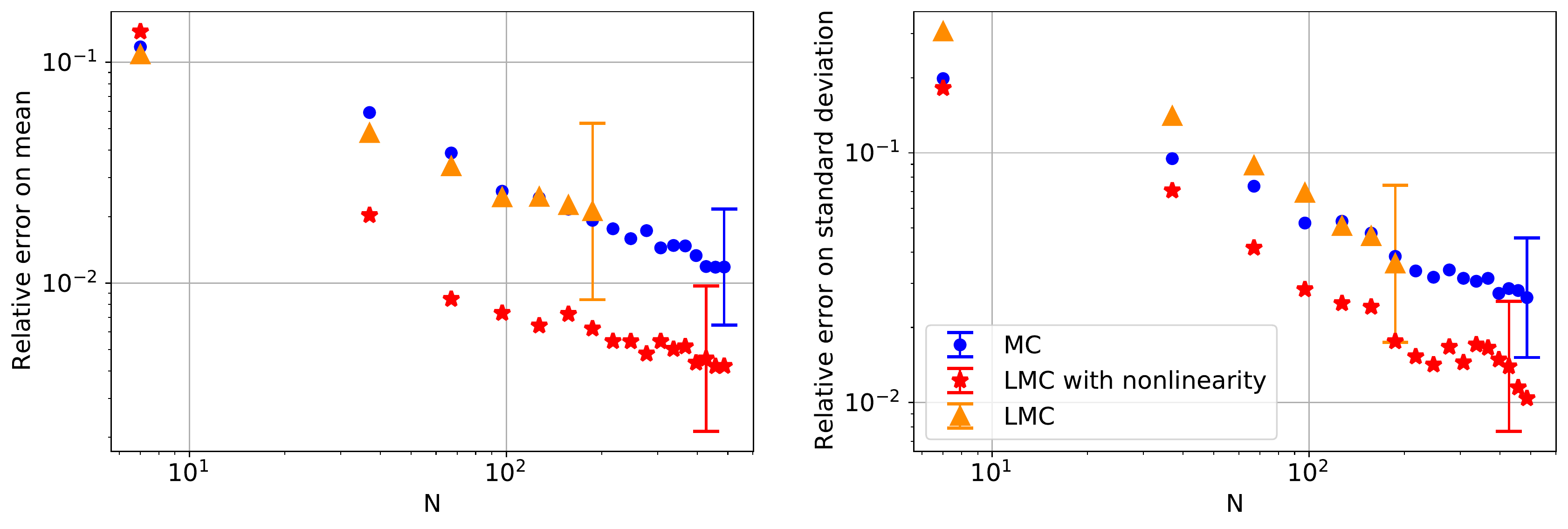}
    \caption{Convergence and relative errors on the estimation of the mean and standard deviation of the Sobol function with input dimension $d=400$, with respect to the computational budget $N$. The markers are the mean of 20 evaluations with each method, and the error bars are the standard deviation of the 20 evaluations. Only a few error bars have been plotted to improve the readability of the plot.}
    \label{fig:sobol_d400_conv}
\end{figure}

Figure \ref{fig:sobol_d400_conv} compares simple MC to LMC with and without the nonlinear transformation. In this case LMC was implemented with $M=10^4$, and the regularisation parameter $\lambda$ was chosen with 5-fold cross-validation. As expected, simple MC and the plain LMC implementation have equal errors. By adding a nonlinear transformation to Lasso, the errors are reduced and LMC becomes more accurate than simple MC.

\subsubsection{Comparison to PCE}
\label{sec:PCE_benchmark}
It is of interest to compare LMC to other common UQ methodologies, such as Polynomial Chaos Expansion (PCE) \cite{sudret_global_2008}. PCE is a well-established technique for UQ and sensitivity analysis, and it is often also benchmarked against the Sobol function \eqref{eq:sobol}. A short description of the PCE method is provided in appendix \ref{app:PCE}.

The PCE implementation used here is the Chaospy library \cite{feinberg_chaospy_2015}. To fit the PCE parameters $\beta_1,\beta_2,...,\beta_P$ we use Lasso regression (see section \ref{sec:Lasso}), and the training set is the same set that is used for the LMC method, a multivariate uniform distribution. Note that there exist other methods for fitting the PCE parameters (see \cite{sudret_global_2008} and \cite{frey_global_2021} for an overview of common methods). Here the Lasso method was chosen for computing the parameters, since the training set is smaller than the number of polynomials, i.e. $N<P$, and as explained in section \ref{sec:Lasso}, Lasso can deal with this problem by finding sparse solutions (see \textit{sparse PCE} \cite{blatman_adaptive_2011, luthen_sparse_2021}). 

\begin{figure}[H]
    \centering
    \includegraphics[width=\textwidth]{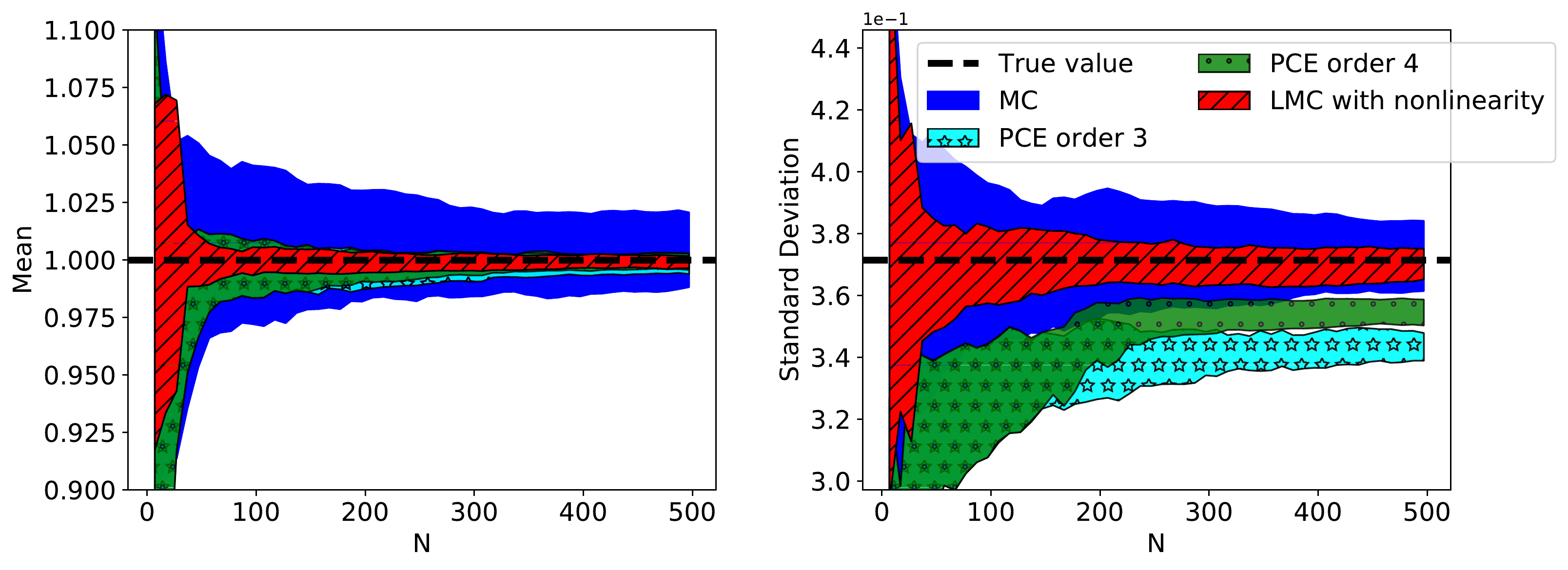}
    \includegraphics[width=\textwidth]{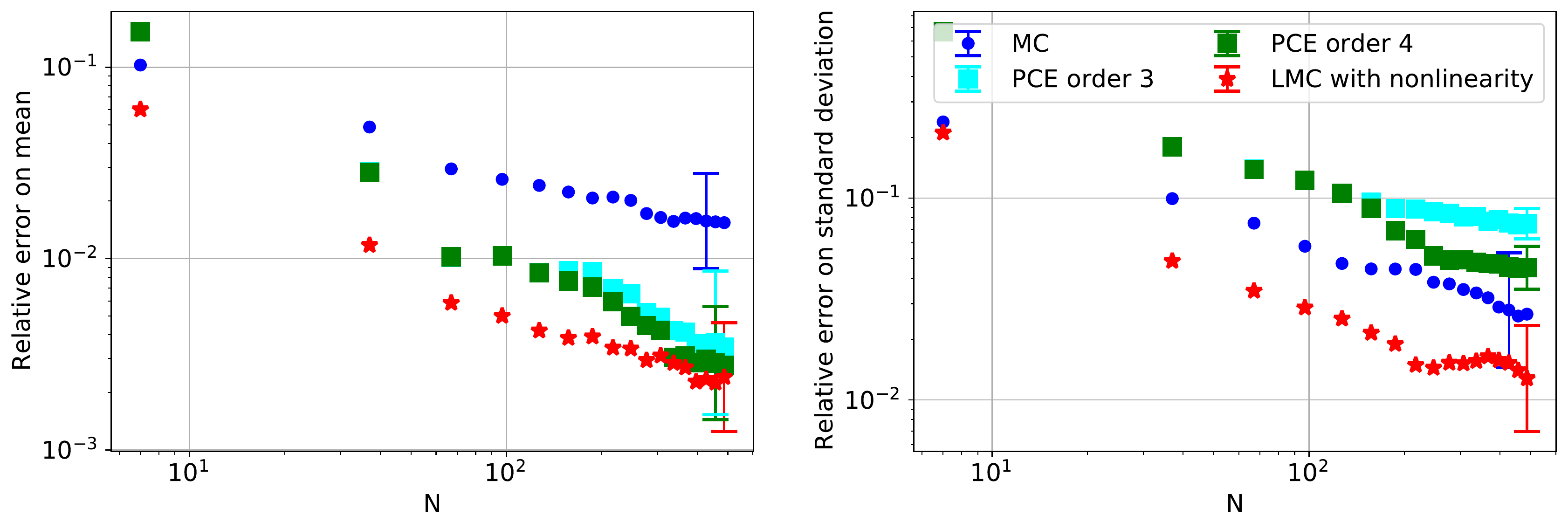}
    \caption{Convergence and relative errors on the estimation of the mean and standard deviation of the Sobol function with input dimension $d=8$, with respect to the size of the training set $N$. The results are the mean of 20 evaluations with each method, and the error bars and coloured-in regions are the standard deviation of the 20 evaluations.}
    \label{fig:sobol_d8_conv}
\end{figure}

Figure \ref{fig:sobol_d8_conv} shows the comparison between MC, LMC with the nonlinear transformation as discussed above, and PCE with orders $p=3,4$. The $p=5$ calculations are not plotted since they looked identical to the $p=4$ runs. \red{In this benchmark the Sobol function of dimension $d=8$ was used, as the Chaospy implementation of PCE became too computationally expensive for larger input dimensions.} Similar to section \ref{sec:linear_benchmark}, three distinct behaviours can be observed: 

\begin{itemize}
    \item simple MC is unbiased but \red{has a large errors},
    \item LMC is unbiased and \red{has smaller errors} than any of the other methods,
    \item PCE, a surrogate-based method, is heavily biased, especially in the estimation of the variance. Furthermore, we note the different behaviour depending on the order of the PCE. By increasing from order $p=3$ to $p=4$,  the number of parameters to be fitted increases from $P=165$ to $P=495$ (see table \ref{tab:P_for_PCE}). As a consequence, the higher order model has a smaller bias when $N>165$.
\end{itemize}


\subsection{FPUT Lattice Problem}
The FPUT lattice problem, proposed by Fermi, Pasta, Ulam, and Tsingou in 1955 \cite{dauxois_fermipastaulam_2005}, models a long chain of $P$ nonlinear oscillators. The FPUT equations of motion, with boundary conditions $x_0(t) = 0,\,x_{P+1}(t) = 1 \,\forall t\in[0,T]$ and 
initial conditions $\dot{\mvec x}(t=0) = \mvec v_0,\,\mvec x(t=0) = \mvec x_0$, can be stated as:

\begin{equation} \label{eq:FPUT}
\ddot{x}_j(t) = \frac{k}{m_j}\left(x_{j+1} + x_{j-1} - 2x_j\right) \left[1 + \alpha(x_{j+1} - x_{j-1})\right], \quad\forall j = 1,2, \ldots ,P.
\end{equation}

\begin{figure}[H]
    \centering
        \includegraphics{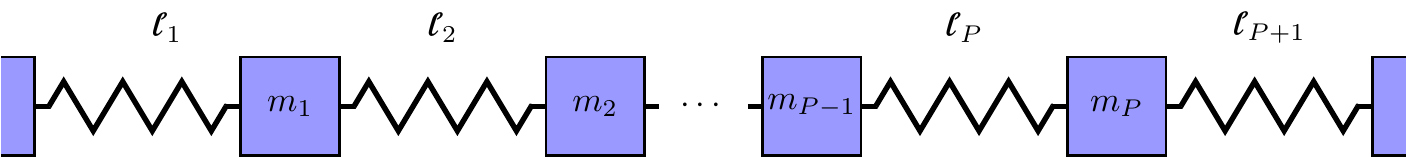}
    \caption{Schematics of the FPUT model.}
    \label{fig:FPUT}
\end{figure}

With $\mvec x(t) = (x_1,x_2,...,x_P)(t)$ and $x_j(t)$ we denote the position of the $j$-th mass in the chain, and $T$ the final time. By defining $\mathbcal{l}_j = x_j-x_{j-1}$ as in figure \ref{fig:FPUT}, and redefining the coupling constant $k^\prime_j\coloneqq\frac{k}{m_j}$, the equations of motion can be rewritten as

\begin{equation*}
    \ddot{x}_j = k^\prime_j\left(\mathbcal{l}_{j+1} - \mathbcal{l}_{j}\right) + \alpha k^\prime_j(\mathbcal{l}^2_{j+1} - \mathbcal{l}^2_{j})\,,\quad\forall j = 1,2,...,P\,,
\end{equation*}

with the boundary and initial conditions from \eqref{eq:FPUT}.

For the numerical experiments, we fix $P=40$, set arbitrary initial conditions

\begin{equation*}
    \begin{cases}
        x_{0,j} &= j\frac{1}{P+1}\,,\\
        v_{0,j} &= \frac{1}{5}\sin\left(3\pi x_j\right)\,,\\
    \end{cases}
    \quad \forall j=1,2,...,P\,,
\end{equation*}

and let the final time be $T=500$ . We now define the UQ problem \red{as estimating the uncertainty of the final kinetic energy $E_K(T)$, given an uncertainty in the spring constants and the nonlinear parameter $\alpha$. This allows for a nonlinear problem of arbitrary input dimension $d=P+1$. The UQ problem can be defined as estimating the mean and variance of the following function}

\begin{equation}
    f\colon \begin{array}[t]{ >{\displaystyle}r >{{}}c<{{}}  >{\displaystyle}l } 
          \mathbb{R}^{P+1} &\to& \mathbb{R} \\ 
          \left(k^\prime_1,k^\prime_2,...,k^\prime_P, \alpha\right) &\mapsto& f(k^\prime_1,k^\prime_2,...,k^\prime_P, \alpha)=E_K(T)\,, 
         \end{array}
    \label{eq:FPUT_UQ}
\end{equation}

where $E_K(T)$ is the kinetic energy of the system at time $T$, calculated as

\begin{equation*}
    E_K(t) = \frac{1}{2}\sum_{j=1}^{N}m_j\dot{x}^2_j(t)\,,
\end{equation*}

and the input arguments of \eqref{eq:FPUT_UQ} are sampled from a normal distribution

\begin{equation}
    \mathcal{N}\left(\left(1,1,...,1,\frac{1}{2}\right), \sigma^2\right)\,,\quad \text{with } \sigma=10^{-3}\,.
    \label{eq:FPUT_init}
\end{equation}

The outputs of \eqref{eq:FPUT_UQ} are obtained by simulating the FPUT model during time $t\in[0,T]$, and the final kinetic energy $E_K(T)$ is computed from the velocities in the final timestep. To simulate the FPUT model we integrate the equations of motion \eqref{eq:FPUT} using the Scipy implementation of RK45 \cite{virtanen_scipy_2020, dormand_family_1980}.

\begin{figure}[H]
    \centering
    \includegraphics[width=\textwidth]{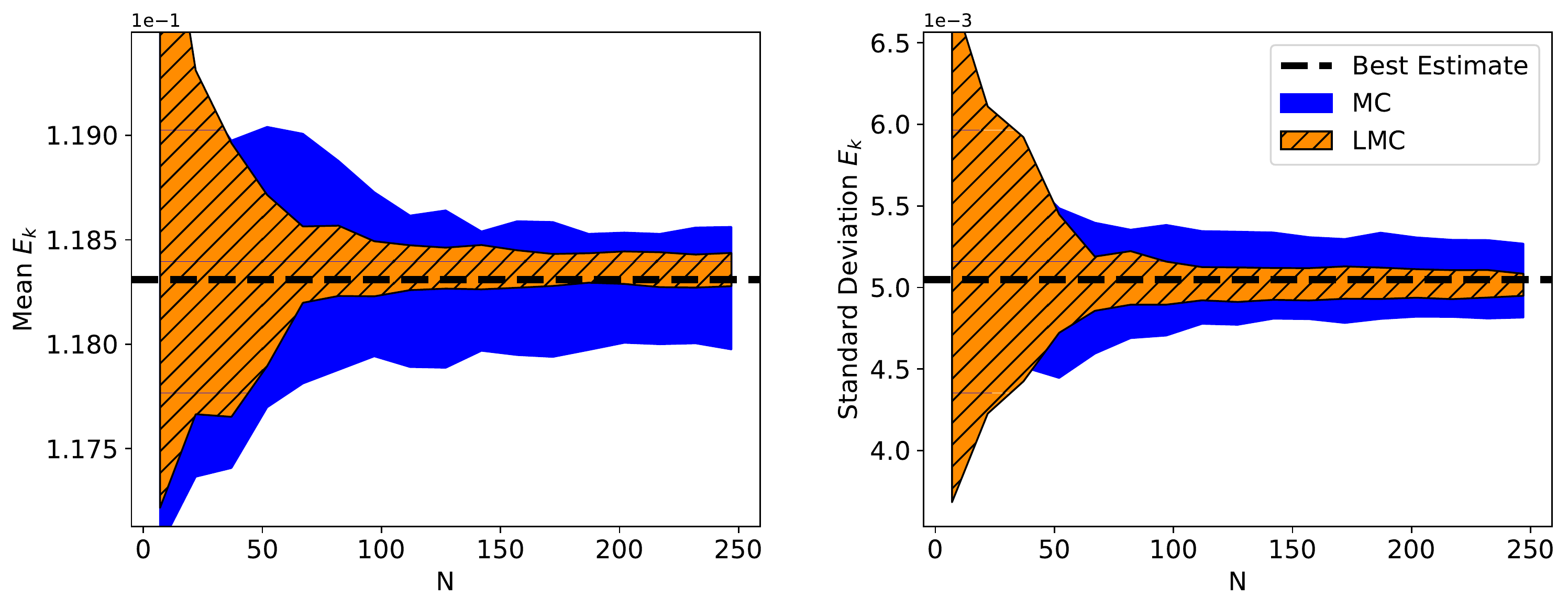}
    \includegraphics[width=\textwidth]{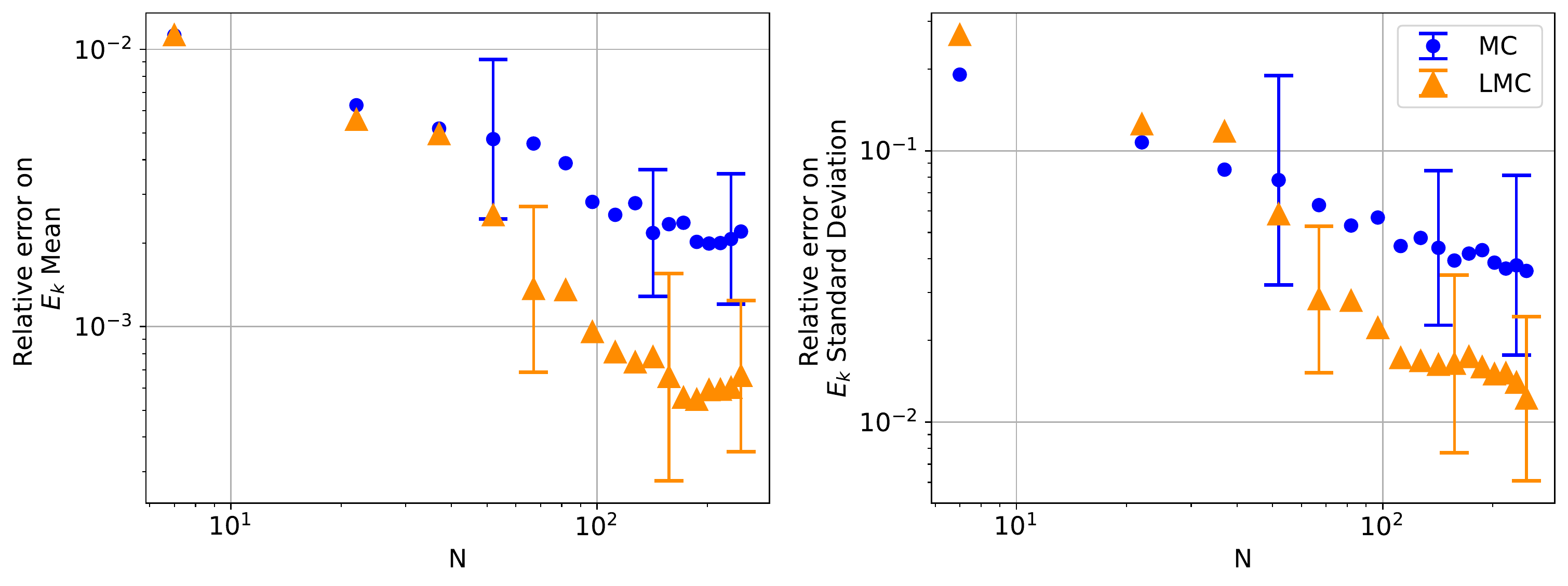}
    \caption{Convergence and relative errors on the final kinetic energy of the FPUT simulation. The results are the mean of 20 evaluations with each method, and the error bars and coloured-in regions are the standard deviation of the 20 evaluations.}
    \label{fig:conv_FPUT}
\end{figure}

Figure \ref{fig:conv_FPUT} shows the convergence of simple MC and LMC, when used on the UQ problem \eqref{eq:FPUT_UQ} with input distribution \eqref{eq:FPUT_init}. For this benchmark the LMC algorithm (alg. \ref{alg:LMC}) was used with $M=10^4$. We note that the authors also carried out this UQ problem with different initial conditions and input distributions, and that in several cases the LMC method did not provide any advantage over simple MC, as the two methods \red{showed similar accuracy}. The distribution \eqref{eq:FPUT_init} used in figure \ref{fig:conv_FPUT} was selected for being a case where LMC is clearly advantageous over simple MC.

\subsection{Nuclear Assembly Burnup Calculations}
\label{sec:nuclear_benchmark}

\begin{figure}[H]
    \centering
    \includegraphics[width = \textwidth]{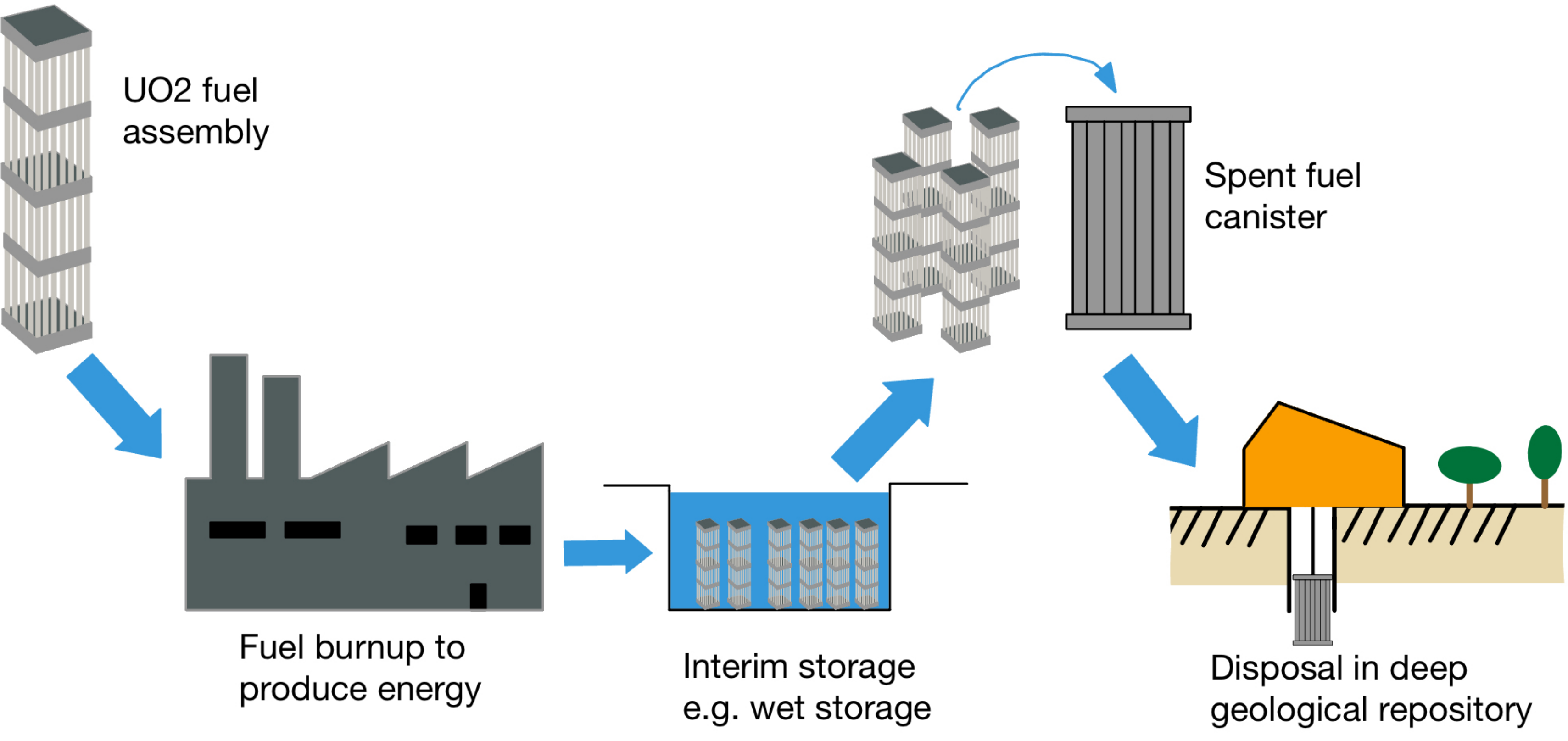}
    \caption{Diagram of a simplified nuclear fuel cycle with final repository and no reprocessing.}
    \label{fig:diagram_FA}
\end{figure}

Spent nuclear fuel (SNF) can be hazardous for tens of thousands of years after its removal from a nuclear reactor, due to decay heat, radiation, and possible criticality excursions. It is therefore essential to accurately calculate quantities such as the decay heat to reduce the risks and costs of interim storage and disposal of SNF (a diagram of the fuel's life cycle is shown in figure \ref{fig:diagram_FA}). Moreover, SNF calculations must account for uncertainties, either with a conservative approach, or with the \textit{best estimate plus uncertainty} (BEPU) approach. In the latter case, UQ is included in the SNF calculations, for example by using Monte Carlo sampling of the input covariances.

The decay heat emitted by spent nuclear fuel is a quantity of interest that needs to be accurately known, to reduce the risks and costs of interim storage and final disposal of the spent fuel. Decay heat and its uncertainty can be numerically estimated with nuclear codes that simulate a nuclear fuel assembly during its irradiation and cooling phases. An example of such a code is CASMO5 \cite{rhodes_casmo-5_2006}, which can be roughly described with

\begin{equation}
    \texttt{CASMO5}:
    \underbrace{
    \begin{pmatrix}
        \text{Fresh fuel parameters} \\
        \text{Irradiation history} \\
        \text{Reactor parameters}\\
        \text{Nuclear Data}
    \end{pmatrix}
    }_{\text{Uncertain Input}}
    \rightarrow
    \begin{pmatrix}
      \text{Decay Heat}\\
      \text{Isotopic Content}\\
      \text{etc...}
    \end{pmatrix}\,.
    \label{eq:CASMO_function}
\end{equation}

To calculate the uncertainty of the decay heat, a UQ process is carried out to propagate the uncertainty of the inputs to the outputs of CASMO5. Generally speaking, nuclear data is a term used to describe the physical quantities (e.g. cross sections) that appear in the equations that CASMO5 solves. These quantities are provided in so-called \textit{nuclear data libraries} (e.g. \cite{plompen_joint_2020, brown_endfb-viii0_2018}) which contain the nominal values of nuclear data and their covariance (uncertainties plus their correlation matrix). Furthermore, in the literature it is common practice to treat the uncertainty of nuclear data separately from other sources of uncertainty \cite{vasiliev_preliminary_2019}, and, for propagating nuclear data uncertainty, simple MC is a widely used and accepted method (e.g. \cite{zwermann_nuclear_2014, leray_methodology_2017, rochman_nuclear_2020}). Nuclear data libraries comprise tens of thousands of parameters, with an associated covariance that propagates to the output of \eqref{eq:CASMO_function}.
In this example, the only input and output of interest are nuclear data and decay heat respectively, and thus the UQ problem can be described with
\begin{eqnarray*}
 f\colon  \mathbb{R}^{15\,557} &\to& \mathbb{R}  \\
\left(\textit{nuclear data}\right) &\mapsto& f\left(\textit{nuclear data}\right) = \text{Decay Heat} 
\end{eqnarray*}
with the input nuclear data sampled from a normal distribution with a covariance matrix $\Sigma$ provided by the nuclear data library. 

The output $f\left(\textit{nuclear data}\right)$ is computed with CASMO5, where the remaining inputs in \eqref{eq:CASMO_function} are kept constant. To generate the input samples, the in-house code developed at PSI, SHARK-X \cite{wieselquist_psi_2013, aures_benchmarking_2017}, is used. With this tool $15~557$ nuclear data values are perturbed according to the nuclear covariance matrix of the ENDF/B-VII.1 library \cite{chadwick_endfb-vii1_2011}, and provided as input to CASMO5. The perturbed values include the scattering elastic and inelastic cross-sections, (n,2n) reactions, neutron induced fission, neutron capture, neutron multiplicity, fission yields, and the fission spectrum, for 19 discrete energy groups. The modelled representative fuel assembly comes from the Ringhals-2 pressurised water reactor in Sweden \cite{sturek_measurements_2006}. It consists of 225 pins of UO$_2$ fuel with $3.095 \%$ enrichment, and is irradiated during four reactor cycles up to a total burnup of 35.7 MWd/kgU. The output quantity corresponds to the decay heat of the representative fuel assembly after 500 days of cooling.

The UQ process needs to be carried out for every representative spent fuel assembly that will be disposed of. If the simple MC approach is used, thousands of CASMO5 simulations are required for each UQ procedure. The computational time required for each simulation ranges from 10 minutes to several hours, and in Switzerland over twelve thousand fuel assemblies \cite{solans_loading_2020} are expected for final disposal. Therefore it is of very high interest to reduce the computational cost of UQ.

\begin{figure}[H]
    \centering
    \includegraphics[width=\textwidth]{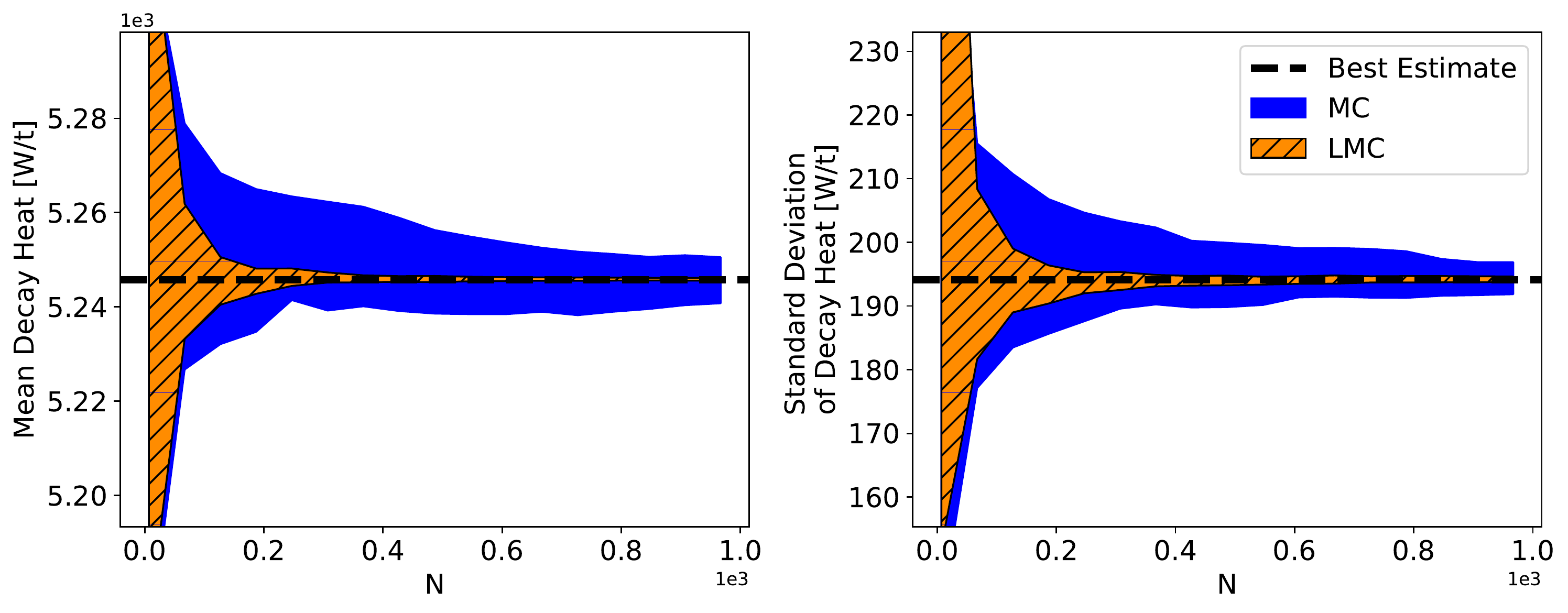}
    \includegraphics[width=\textwidth]{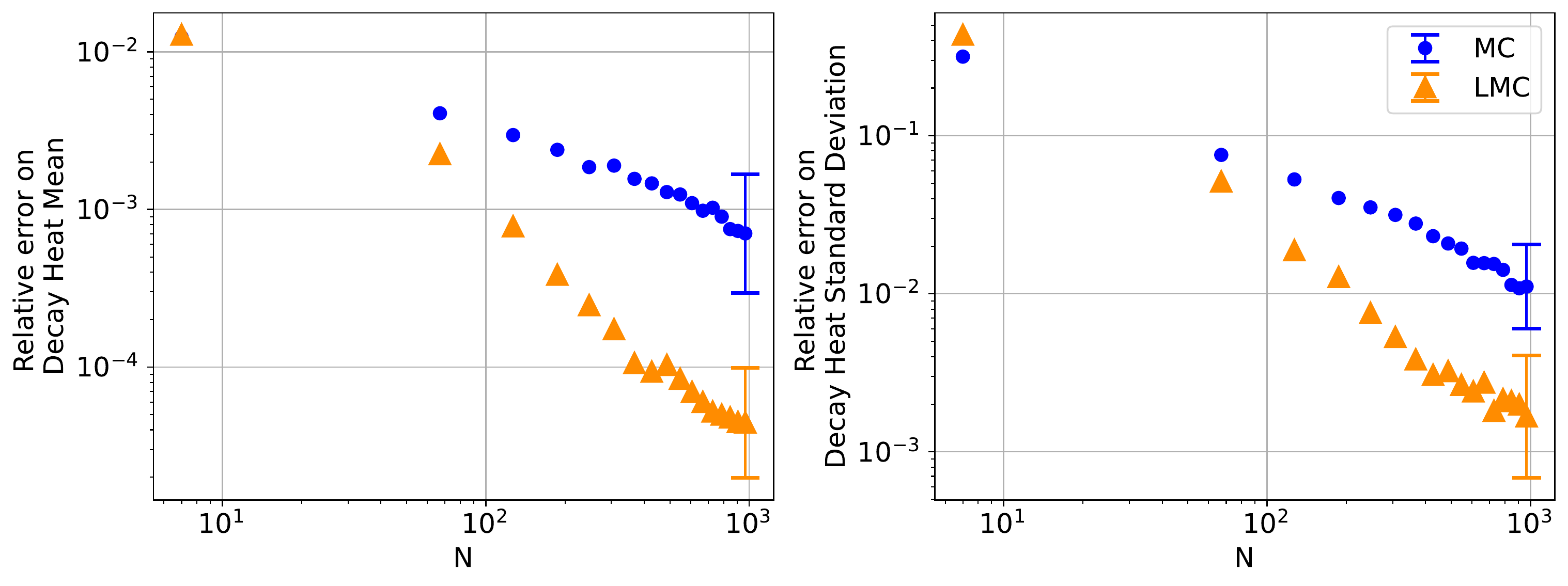}
    \caption{Convergence and relative errors on the estimation of the mean and standard deviation of the decay heat at 500 days of cooling. The results are the mean of 20 evaluations with each method, and the error bars and coloured-in regions are the standard deviation of the 20 evaluations. Only the final error bars were included to improve readability of the plot.}
    \label{fig:conv_DH}
\end{figure}

Figure \ref{fig:conv_DH} shows the estimation of the mean and variance of the decay heat, expressed in Watts per tonne of heavy metal, where it can be seen \red{that LMC is clearly more accurate than simple MC}. At $N=1000$, the simple MC method has a mean relative error of $10^{-3}$ and $10^{-2}$ for the mean and standard deviation respectively, whereas the LMC method requires less than $N<200$ to obtain the same relative errors. This represents a reduction in computational costs by more than a factor of 5. For this benchmark the LMC algorithm (alg. \ref{alg:LMC}) was used with $M=6000$.

\section{Conclusion \& Outlook}
High-dimensional UQ problems are ubiquitous in science, and \red{usually} approached with surrogate-based or simple MC methods. These methods become computationally expensive in high-dimensional settings, and in such problems the proposed LMC can play an important role. Moreover, given a set of input-output data that has been previously used for UQ, LMC can be readily applied without any extra evaluations of $f$, and immediately provide a better estimate of the UQ. 

From the theory point of view, further research is needed to gain a deeper understanding on the choice of regularisation parameter $\lambda$, and on the conditions imposed on $f$ to ensure \red{good accuracy. Indeed, despite numerical examples showing that LMC has smaller errors than other surrogate-based, MC, and multifidelity methods, the theory is still lacking in the non-asymptotic case $N<\infty$}.

As further possible investigations, it is worth noting that LMC could in principle be used with other parametric models a part from Lasso. More complex models have the potential to better fit the data and further reduce the number of computationally intensive function evaluations. The only requirement is for the model to use some form of regularisation, such as the $L^1$-norm, which provides sparsity, a form of model selection\red{, and a small model variance. }

\section*{Acknowledgements}
The authors would like to thank Swissnuclear for partially sponsoring
this work.

\Urlmuskip=0mu plus 1mu\relax  
\bibliographystyle{unsrt}
\bibliography{LMC}

\appendix
\section{Proofs}
\label{app:proofs}

\begin{proof}[Proof of Theorem \eqref{thm:LMC_equal}]

\begin{equation*}
    \mathcal{M}_{N,M} = \frac{1}{S}\sum_{{\smalls}=1}^S \left(\mu\ntr{n}_{N/S.M}\right)_{\smalls} = \mu\ntr{n}_{M} + \frac{1}{S}\sum_{{\smalls}=1}^S \left(\mu_{N/S} - \mu\ntr{n}_{N/S}\right)_{\smalls} = \mu\ntr{n}_M + \mu_N - \mu\ntr{n}_N = \mu\ntr{n}_{N,M}.
\end{equation*}

Therefore $\mathcal{M}_{N,M}=\mu\ntr{n}_{N,M}$ and they have the same mean squared error.

\begin{align*}
    \MSE&\left(\Sigma^2_{N,M}, \Var[f]\right) = \MSE\left(\frac{1}{S}\sum_{\smalls=1}^S\left({\sigma^2}\ntr{n}_{N/S,M}\right)_{\smalls} , \Var[f]\right) \\
    &= \E\left[\left({\s2}\ntr{n}_{M} - \Var[f] + \frac{1}{S}\sum_{\smalls=1}^S\left({\s2}_{N/S} - {\s2}\ntr{n}_{N/S}\right)_{\smalls}\right)^2\right]\\
    &=\Var\left[{\s2}\ntr{n}_{M}\right] + \Var\left[\frac{1}{S}\sum_{\smalls=1}^S\left({\s2}_{N/S} - {\s2}\ntr{n}_{N/S}\right)_{\smalls}\right] +  2\underbrace{\Cov\left[{\s2}\ntr{n}_M,  \frac{1}{S}\sum_{\smalls=1}^S\left({\s2}_{N/S} - {\s2}\ntr{n}_{N/S}\right)_{\smalls}\right]}_{=0}\\
    &=\Var\left[{\s2}\ntr{n}_{M}\right] + \frac{1}{S^2}\sum_{\smalls=1}^S\Var\left[\left({\s2}_{N/S} - {\s2}\ntr{n}_{N/S}\right)_{\smalls}\right] +  \frac{1}{S^2}\sum_{\substack{\smalls,t=1\\\smalls\neq t}}^S\underbrace{\Cov\left[\left({\s2}_{N/S} - {\s2}\ntr{n}_{N/S}\right)_{\smalls},\left({\s2}_{N/S} - {\s2}\ntr{n}_{N/S}\right)_t\right]}_{=0}\\
    &=\Var\left[{\s2}\ntr{n}_{M}\right] + \frac{S}{S^2}\Var\left[{\s2}_{N/S} - {\s2}\ntr{n}_{N/S}\right] = \Var\left[{\s2}\ntr{n}_{M}\right] + \frac{1}{S}\Var\left[{\s2}_{N/S} - {\s2}\ntr{n}_{N/S}\right]\,.\\
\end{align*}

The last equation is similar to the MSE of a two-level estimator ${\s2}\ntr{n}_{N/S,M}$, which is known from eq. \eqref{eq:LMC_var_MSE}. By replacing the MSE in this equation we have

\begin{align*}
    \Var&\left[{\s2}\ntr{n}_{M}\right] + \frac{1}{S}\Var\left[{\s2}_{N/S} - {\s2}\ntr{n}_{N/S}\right] = \Var\left[{\s2}\ntr{n}_{M}\right]\\
    &+\frac{1}{S}\frac{1}{\frac{N}{S}}\left(m_{2,2}\left[f+\surr{n}, f-\surr{n}\right] + \frac{1}{\frac{N}{S}-1}\Var[f+\surr{n}]\Var[f-\surr{n}] - \frac{\frac{N}{S}-2}{\frac{N}{S}-1}\left(\Var[f]-\Var[\surr{n}]\right)^2\right)\\
    &= \Var\left[{\s2}\ntr{n}_{M}\right] + \Var\left[{\s2}_N - {\s2}\ntr{n}_N\right] +  \frac{1}{N}\left(\frac{1}{\frac{N}{S}-1} - \frac{1}{N-1}\right)\Var[f+\surr{n}]\Var[f-\surr{n}]\\
    & + \frac{1}{N}\left(\frac{1}{\frac{N}{S}-1} - \frac{1}{N-1}\right)\left(\Var[f]-\Var[\surr{n}]\right)^2\\
    &=\MSE\left({\s2}\ntr{n}_{N,M} , \Var[f]\right) + \mathcal{O}\left(N^{-2}\right)\,.\\
\end{align*}

\end{proof}

\begin{proof}[Proof of Lemma \eqref{lem:lambda_max}]
The minimum of the loss function is given by the stationary point satisfying

\begin{equation*}
    \frac{\partial}{\partial \beta_{k}}\mathcal{L}(\mvec\beta) = 0\quad\forall\, k=1,2,...,d\,,
\end{equation*}

which is a minimum since both terms in \eqref{eq:lasso_loss} are convex. If we write this equation explicitly, using sub-derivatives since $|\beta_k|$ is nondifferentiable at $\beta_k=0$, one finds

\begin{align}
        \nonumber&-\sum_{i=1}^{n}x_{ik}\left(f(\mvec x_i) - \mvec\beta\cdot\mvec x_i \right) + \lambda\frac{\partial}{\partial \beta_{k}}|\beta_k| = 0\\
        \nonumber&\iff \sum_{i=1}^{n} f(\mvec x_i)x_{ik} - \sum_{\substack{l=1\\l\neq k}}^d\beta_{l}\sum_{i=1}^{n}x_{il}x_{ik} = \lambda\frac{\partial}{\partial \beta_{k}}|\beta_k| + \beta_k\sum_{i=1}^{n}x_{ik}^2\\
        &\iff \begin{cases}
            \frac{1}{\sum_{i=1}^{n}x_{ik}^2}\left(\sum_{i=1}^{n} f(\mvec x_i)x_{ik} - \lambda\sgn(\beta_k) - \sum_{\substack{l=1\\l\neq k}}^d\beta_{l}\sum_{i=1}^{n}x_{il}x_{ik}\right) = \beta_k&\quad\text{if}\,\beta_k\neq 0\,,\\
            &\\
            \sum_{i=1}^{n} f(\mvec x_i)x_{ik} - \sum_{\substack{l=1\\l\neq k}}^d\beta_{l}\sum_{i=1}^{n}x_{il}x_{ik}\in[-\lambda, \lambda]&\quad\text{if}\,\beta_k= 0\,.\\
        \end{cases}
    \label{eq:lemma}
\end{align}

If $\mvec\beta=(0,0,...,0)$ is a solution to \eqref{eq:lemma}, then $\lambda\geq\lambda_{max}$, with $\lambda_{max}$ defined in \eqref{eq:lambda_max}. Moreover, this solution is unique since $\{x_{ik}\}$ comes from a continuous probability distribution \cite{tibshirani_lasso_2013}. Hence, since $\mvec\beta=\mvec 0$ is a solution if $\lambda\geq\lambda_{max}$, it is the only solution and thus  $\lambda\geq\lambda_{max}\Rightarrow\mvec\beta=\mvec 0$.
\end{proof}

\begin{proof}[Proof of Theorem \eqref{thm:Lasso_Mean} (a)] A weight vector $\mvec\beta$ that minimises the Lasso loss \eqref{eq:lasso_loss} must satisfy \eqref{eq:lemma}. If \eqref{eq:lemma} is multiplied on both sides by $\frac{\beta_k}{n-1}$ we have

\begin{equation}
    \beta_k\frac{1}{n-1}\sum_{i=1}^{n} f(\mvec x_i)x_{ik} =
    \beta_k\sum_{l=1}^d\beta_{l}\frac{1}{n-1}\sum_{i=1}^{n}x_{il}x_{ik}  + \frac{\lambda}{n-1}|\beta_k|\quad\forall\, k=1,2,...,d\,.
    \label{eq:proofA_step1}
\end{equation}

In this equation, the left hand side and the first term of the right hand side contain the unbiased estimator for the covariance, which reads

\begin{equation*}
    \lim_{n\to\infty}\frac{1}{n-1}\sum_{i=1}^{n} \left(z_i - \frac{1}{n}\sum_{j=1}^{n}z_j\right)\left(y_i - \frac{1}{n}\sum_{j=1}^{n}y_j\right) = \Cov[Z,Y],
\end{equation*}

for some random variables $Z$ and $Y$. Then, in the limit $n\rightarrow\infty$, equation \eqref{eq:proofA_step1} becomes

\begin{equation}
    \beta_{k}\Cov[f,X_k]=\sum_{l=1}^d\beta_{k}\beta_{l}\Cov[X_l,X_k] + \underbrace{\frac{\lambda}{n-1}}_{\rightarrow 0}|\beta_{k}|.
    \label{eq:cov_identity}
\end{equation}

Now, by inserting this expression into assumption \eqref{eq:assumption_a}, the assumption is shown to be satisfied:

\begin{equation}
    \begin{split}
        \Var[f-\surr{n}] &= \Var[f] + \Var[\surr{n}] - 2\Cov(f,\surr{n}) = \Var[f] + \Var[\surr{n}] - 2\sum_{k=1}^d\beta_{k}\Cov(f,X_k)\\
        &= \Var[f] + \Var[\surr{n}] - 2\sum_{k=1}^d\sum_{l=1}^d\beta_{k}\beta_{l}\Cov(X_k,X_l)\\
        &= \Var[f] + \Var[\surr{n}] - 2\Var[\surr{n}] = \Var[f] - \Var[\surr{n}] \leq \Var[f].
    \end{split}
    \label{eq:proofA_step3}
\end{equation}
\end{proof}

\begin{proof}[Proof of Theorem \eqref{thm:Lasso_Mean} (b)] For the sake of simplifying notation we define the function

\begin{equation*}
    S(\mvec\beta)=\Var[f-\surr{n}] - \Var[f]\,,
\end{equation*}

such that assumption \eqref{eq:assumption_a} is satisfied if and only if $S(\mvec\beta)\leq 0$. This function can be rewritten as

\begin{equation}
    S(\mvec\beta) = \Var[\surr{n}] - 2\Cov[f,\surr{n}] = \sum_{k=1}^d\sum_{l=1}^d\beta_k\beta_l\Cov[X_k,X_l] - 2\sum_{k=1}^d\beta_k\Cov[f,X_k]\,.
    \label{eq:S}
\end{equation}

Its gradient and Hessian are given by

\begin{align}
    \frac{\partial S}{\partial \beta_m}(\mvec\beta) &= 2\left(\sum_{k=1}^d\beta_k\Cov[X_k,X_m] -\Cov[f,X_m] \right)\,,\label{eq:S_first_deriv}\\
    \frac{\partial^2 S}{\partial\beta_m \partial\beta_n}(\mvec\beta) &= 2\Cov[X_m,X_n]=2\,\Sigma_{mn}\label{eq:S_second_deriv}\,.
\end{align}

Since the Hessian \eqref{eq:S_second_deriv} is proportional to the covariance matrix $\Sigma\succ 0$, it must be positive definite, and therefore $S(\mvec\beta)$ is strictly convex, and thus any local minimum is the global minimum. From \eqref{eq:S_first_deriv}, the vector $\mvec\beta^*$ that minimises $S(\mvec\beta)$ satisfies

\begin{equation*}
    \Cov[f,X_m] = \sum_{k=1}^d\beta^*_k\Cov[X_k,X_m]\,,
\end{equation*}

and it is clear from \eqref{eq:proofA_step3}, that $S(\mvec\beta^*)\leq 0$. Therefore $S$ has a nonpositive global minimum. From the convexity, it follows that there exists a convex domain $\Omega\subset\mathbb{R}^d$, such that

\begin{equation*}
    S(\mvec\beta)\leq 0\iff\mvec\beta\in\Omega\,.
\end{equation*}

Additionally we have that the null vector $\mvec0=(0,0,...,0)\in\Omega$, since $S(\mvec 0)=0$.

Now, let $\mvec\beta(\lambda)$ be the vector that minimises the loss function \eqref{eq:lasso_loss}, for a given $\lambda>0$. The regularisation path $\left\{\mvec\beta(\lambda) \,|\, \lambda>0\right\}$ is continuous and piecewise linear \cite{efron_least_2004, rosset_piecewise_2007, mairal_complexity_2012}. Moreover, for $\lambda\rightarrow0^+$ we have $\mvec\beta(\lambda)\rightarrow\mvec\beta^{\text{OLS}}$, and if $\lambda\geq\lambda_{max}$, with $\lambda_{max}$ as in \eqref{eq:lambda_max}, then $\mvec\beta(\lambda)=\mvec 0$. Therefore $\mvec\beta(\lambda)$ follows a continuous path from $\mvec\beta^{\text{OLS}}$ to $\mvec 0$ as $\lambda$ increases. In general, the start of the path is not in $\Omega$, since for a finite training set with $n<\infty$ it is possible that $\beta^{\text{OLS}}\notin \Omega$. However, the path ends in $\Omega$ since $\mvec0\in\Omega$. Therefore, there exists at least one point $\lambda^*$ with $0<\lambda^*\leq\lambda_{max}$, such that $\mvec\beta(\lambda^*)\in\Omega$. Thus, $S(\mvec\beta(\lambda^*))\leq 0$.

\end{proof}

\begin{remark}
A schematic of a the two-dimensional projection of the domain $\Omega$ is shown in figure \ref{fig:omega_mean}. Despite that $\mvec\beta^{\text{OLS}}\notin\Omega$, the Lasso regularisation path must have some points in $\Omega$, since it is a continuous path that ends at $\mvec 0$.

\begin{figure}[H]
    \centering
    \begin{tikzpicture}[scale=1.0]
        \draw[->] (-1, 0) -- (4, 0) node[above] {$\beta_k$};
        \draw[->] (0, -1) -- (0, 2) node[above] {$\beta_m$};
        \fill (3,1.5)  circle[radius=2pt] node[above] {$\mvec\beta^*$};
        \fill (2.8,-.5)  circle[radius=2pt] node[right] {$\mvec\beta^{\text{OLS}}$};
        
        \draw[blue,dashed,line width=0.8mm,] (2.8,-.5) -- (2,0);
        \draw[blue,dashed,line width=0.8mm,] (2,0) -- (0,0);
        \draw[green,dashed,line width=0.8mm,] (2.8,-.5) -- (2.1,1);
        \draw[green,dashed,line width=0.8mm,] (2.1,1) -- (.2,.5);
        \draw[green,dashed,line width=0.8mm,] (.2,.5) -- (0,0);
        \draw[olive,dashed,line width=0.8mm,] (2.8,-.5) -- (.5,-1);
        \draw[olive,dashed,line width=0.8mm,] (.5,-1) -- (0,0);
        
        \begin{scope}[shift={(0,0))},x={(3.6,1.8)},y={($(0,0)!1!90:(3.6,1.8)$)}]
            \draw[fill=red,fill opacity=0.2, draw=none] (.5,0) ellipse (.5 and .07);
        \end{scope}
        \node[text width=0cm] at (1,1.2) {\red{$\Omega$}};
    \end{tikzpicture}
    \caption{Example 2D projection of the domain $\Omega$. The dashed lines represent three possible regularisation paths.}
    \label{fig:omega_mean}
\end{figure}
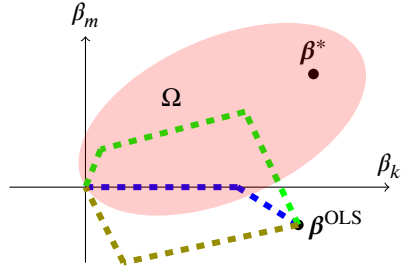

\end{remark}

\begin{proof}[Proof of Lemma \eqref{lem:Lasso_fail}] The left hand side of assumption \eqref{eq:assumption_b} can be rewritten as

\begin{equation*}
    \begin{split}
        &m_{2,2}\left[f+\surr{n}, f-\surr{n}\right] + \frac{1}{N-1}\Var[f+\surr{n}]\Var[f-\surr{n}] - \frac{N-2}{N-1}\left(\Var[f]-\Var[\surr{n}]\right)^2\\
        &=m_4[f] + m_4[\surr{n}] - 2m_{2,2}[f,\surr{n}] + \frac{1}{N-1}\left({\Var}^2[f] + {\Var}^2[\surr{n}] + 2\Var[f]\Var[\surr{n}] - 4{\Cov}^2[f,\surr{n}]\right)\\
        &\quad - \frac{N-2}{N-1}\left({\Var}^2[f] + {\Var}^2[\surr{n}] - 2\Var[f]\Var[\surr{n}]\right)\\
        &= m_4[f] - \frac{N-3}{N-1}{\Var}^2[f] \\
        &\quad+ m_4[\surr{n}] - \frac{N-3}{N-1}{\Var}^2[\surr{n}] - 2\left(m_{2,2}[f,\surr{n}] - \Var[f]\Var[\surr{n}]\right) - \frac{4}{N-1}{\Cov}^2[f,\surr{n}]\,.\\
    \end{split}
\end{equation*}

By comparing this expression to the right hand side of \eqref{eq:assumption_b}, the assumption can be rewritten as

\begin{align}
    \nonumber m_4&[\surr{n}] - \frac{N-3}{N-1}{\Var}^2[\surr{n}] - 2\left(m_{2,2}[f,\surr{n}] - \Var[f]\Var[\surr{n}]\right) \\
    \nonumber &- \frac{4}{N-1}{\Cov}^2[f,\surr{n}]\leq0\quad\forall\,N>2\,,\\
    &\iff \Var[{\surr{n}}^2] - 2\Cov[f^2, {\surr{n}}^2] + \frac{2}{N-1}\left({\Var}^2[\surr{n}] - 2{\Cov}^2[f,\surr{n}]\right)\leq 0\quad\forall\,N>2\,,
    \label{eq:assumption_b_simplified}
\end{align}

It suffices to show that \eqref{eq:assumption_b} is not satisfied in the asymptotic limit $n\rightarrow\infty$. From eq. \eqref{eq:cov_identity} we conclude that, in the asymptotic limit, we have

\begin{equation*}
    \Cov[f,\surr{n}] = \Var[\surr{n}]\,,
\end{equation*}

and by injecting this identity into \eqref{eq:assumption_b_simplified} we have the simpler inequality

\begin{equation}
    \Var[{\surr{n}}^2] - 2\Cov[f^2, {\surr{n}}^2] - \frac{2}{N-1}{\Var}^2[\surr{n}]\leq 0\quad\forall\,N>2\,.
    \label{eq:assumption_b_asymptotic}
\end{equation}

Since this condition applies to all $N>2$, we can simply show that it is not satisfied when $N\rightarrow\infty$. Then assumption \eqref{eq:assumption_b_asymptotic} becomes

\begin{equation}
    \Var[{\surr{n}}^2] - 2\Cov[f^2, {\surr{n}}^2] \leq 0\,.
    \label{eq:assumption_b_doubleasymptotic}
\end{equation}

To prove that \eqref{eq:assumption_b_doubleasymptotic} is in general not satisfied, we provide a one-dimensional example $f:\mathbb{R}\rightarrow\mathbb{R}$ that does not satisfy \eqref{eq:assumption_b_doubleasymptotic}. The example is the following: let the input be uniform $X\sim U[-1,1]$, and $f(x)=\sin(\pi x)$. Then in the limit $n\rightarrow\infty$ we have $\surr{n}(x) = \beta x$ with $\beta>0$ and

\begin{equation*}
    \Cov[f^2,{\surr{n}}^2] = \beta^2\Cov[f^2, X^2]=  -\beta^2\frac{1}{4\pi^2}<0\,,
\end{equation*}

and thus \eqref{eq:assumption_b_doubleasymptotic} is not satisfied.

\end{proof}

\begin{proof}[Proof of Theorem \eqref{thm:Lasso_Variance} (a)] 
Start with the asymptotic expression \eqref{eq:assumption_b_asymptotic}, which assumes $n\rightarrow\infty$. Its last term is negative, and therefore

\begin{equation*}
    \Var[{\surr{n}}^2] - 2\Cov[f^2, {\surr{n}}^2] - \underbrace{\frac{2}{N-1}{\Var}^2[{\surr{n}}]}_{\geq 0}\leq 0\quad\Leftarrow\quad\Var[{\surr{n}}^2] - 2\Cov[f^2, {\surr{n}}^2] \leq 0\,.
\end{equation*}

We can then simply show that the the right hand inequality is true. Since the explicit expressions for $f$ and ${\surr{n}}$ are known, we can write

\begingroup
\allowdisplaybreaks
\begin{align*}
   &\Var[{\surr{n}}^2] - 2\Cov[f^2, {\surr{n}}^2] \\
   &= \E\left[\sum_{k,l,m,p=1}^d \beta_k\beta_l\beta_m\beta_p\,X_kX_lX_mX_p\right] - \sum_{k,l=1}^d \beta_k^2\beta_l^2\\
   &\quad\quad - 2\Cov\left[(\mvec\alpha\cdot\mvec X + \Epsilon)^2,(\mvec\beta\cdot\mvec X)^2\right] \\
   &= \sum_{k=1}^d\beta_k^4\left(\E[X_k^4] - 1\right) + \sum_{\substack{k,l=1\\k\neq l}}^d \beta_k^2\beta_l^2\left(3\E[X_k^2X_l^2]-1\right)\\
   &\quad\quad - 2\left(\Cov[(\mvec\alpha\cdot\mvec X)^2,(\mvec\beta\cdot\mvec X)^2] + \Cov[{\Epsilon}^2, (\mvec\beta\cdot\mvec X)^2] + 2\Cov[\Epsilon(\mvec\alpha\cdot\mvec X), (\mvec\alpha\cdot\mvec X)^2]\right) \\
   &=\sum_{k=1}^d\beta_k^4\Var[X
   _k^2] + 2\sum_{\substack{k,l=1\\k\neq l}}^d\beta^2_k\beta^2_l\\
   &\quad\quad - 2\left(\sum_{k,l,m,p=1}^d\alpha_k\alpha_l\beta_m\beta_p\Cov[X_kX_l,X_mX_p] + \sum_{k,l=1}^d\beta_k\beta_l\underbrace{\Cov[{\Epsilon}^2, X_kX_l]}_{=0} + 2\sum_{k,l,m=1}^d\alpha_k\beta_l\beta_m\underbrace{\Cov[\Epsilon X_k, X_lX_m]}_{=0}\right)\\
   &=\sum_{k=1}^d\beta_k^4\Var[X
   _k^2] + 2\sum_{\substack{k,l=1\\k\neq l}}^d\beta^2_k\beta^2_l - 2\left( \sum_{k=1}^d\alpha^2_k\beta^2_k\Var[X^2_k] + \sum_{\substack{k,l=1\\k\neq l}}^d\alpha_k^2\beta_l^2\underbrace{\Cov[X_k^2,X_l^2]}_{=0} + 2\sum_{\substack{k,l=1\\k\neq l}}^d\alpha_k\alpha_l\beta_k\beta_l\underbrace{\Cov[X_kX_l,X_kX_l]}_{=\Var[X_k]\Var[X_l]=1}\right)\\
   &= \sum_{k=1}^d\left(\beta_k^4 - 2\alpha_k^2\beta_k^2\right)\Var[X
   _k^2] + 2\sum_{\substack{k,l=1\\k\neq l}}^d\left(\beta^2_k\beta^2_l - 2\alpha_k\beta_k\alpha_l\beta_l\right)\,.\\
\end{align*}
\endgroup

From \eqref{eq:cov_identity}, in the asymptotic limit $\mvec\beta$ satisfies

\begin{equation*}
    \alpha_k = \Cov[f,X_k] = \sum_{l=1}^d\beta_l\Cov[X_l,X_k] = \underbrace{\Var[X_k]}_{=1}\beta_k = \beta_k\,,\quad \forall k=1,2,...,d\,,
\end{equation*}

and hence 

\begin{equation*}
    \sum_{k=1}^d\left(\beta_k^4 - 2\alpha_k^2\beta_k^2\right)\Var[X
   _k^2] + 2\sum_{\substack{k,l=1\\k\neq l}}^d\left(\beta^2_k\beta^2_l - 2\alpha_k\beta_k\alpha_l\beta_l\right) = -\sum_{k=1}^d\alpha^4_k\Var[X
   _k^2] - 2\sum_{\substack{k,l=1\\k\neq l}}^d\alpha^2_k\alpha^2_l
   \leq 0\,.
\end{equation*}

\end{proof}

\begin{proof}[Proof of Theorem \eqref{thm:Lasso_Variance} (b)] Based on inequality\eqref{eq:assumption_b_simplified}, define the function

\begin{equation*}
    S(\mvec\beta) = \Var[{\surr{n}}^2] - 2\Cov[f^2, {\surr{n}}^2] + \frac{2}{N-1}\left({\Var}^2[{\surr{n}}] - 2{\Cov}^2[f,{\surr{n}}]\right)\,,
\end{equation*}

with $N>2$ an integer, such that \eqref{eq:assumption_b} is satisfied if and only if $S(\mvec\beta)\leq 0$. This function can be rewritten as

\begin{align*}
    S(\mvec\beta) &= \sum_{k=1}^d\left(\beta_k^4 - 2\alpha_k^2\beta_k^2\right)\Var[X
   _k^2] + 2\sum_{\substack{k,l=1\\k\neq l}}^d\left(\beta^2_k\beta^2_l - 2\alpha_k\beta_k\alpha_l\beta_l\right) \\
   &+\frac{2}{N-1}\sum_{k,l=1}^d\left(\beta^2_k\beta^2_l - 2\alpha_k\alpha_l\beta_k\beta_l\right)\\
   &= \sum_{k=1}^d\left(\beta_k^4 - 2\alpha_k^2\beta_k^2\right)\left(\Var[X
   _k^2] + \frac{2}{N-1}\right) + 2\frac{N}{N-1}\sum_{\substack{k,l=1\\k\neq l}}^d\left(\beta^2_k\beta^2_l - 2\alpha_k\alpha_l\beta_k\beta_l\right)\,.\\
\end{align*}

Define the domain $\Omega$ as

\begin{equation*}
    \Omega = \left\{\mvec\beta \,\big|\, 0\leq|\beta_k|\leq|\alpha_k|\,\&\, \sgn(\beta_k) = \sgn(\alpha_k)\quad\forall k=1,2,...,d\right\}\,.
\end{equation*}

Then

\begin{equation*}
    \mvec\beta\in\Omega\Rightarrow S(\mvec\beta)\leq0\,.
\end{equation*}

Now, let $\mvec\beta(\lambda)$ be the vector that minimises the loss function \eqref{eq:lasso_loss}, for a given $\lambda>0$. The regularisation path $\left\{\mvec\beta(\lambda) \,|\, \lambda>0\right\}$ is continuous and piecewise linear \cite{efron_least_2004, rosset_piecewise_2007, mairal_complexity_2012}. Moreover, for $\lambda\rightarrow0^+$ we have $\mvec\beta(\lambda)\rightarrow\mvec\beta^{\text{OLS}}$, and if $\lambda\geq\lambda_{max}$, with $\lambda_{max}$ as in \eqref{eq:lambda_max}, then $\mvec\beta(\lambda)=\mvec 0$. Therefore $\mvec\beta(\lambda)$ follows a continuous path from $\mvec\beta^{\text{OLS}}$ to $\mvec 0$ as $\lambda$ increases. In general, the start of the path is not in $\Omega$, since for a finite training set with $N_{tr}<\infty$ it is possible that $\beta^{\text{OLS}}\notin \Omega$. However, the path ends in $\Omega$ since $\mvec0\in\Omega$. Therefore, there exists at least one point $\lambda^*$ with $0<\lambda^*\leq\lambda_{max}$, such that $\mvec\beta(\lambda^*)\in\Omega$. Thus, $S(\mvec\beta(\lambda^*))\leq 0$.
\end{proof}

\begin{remark}
A schematic of a the two-dimensional projection of the domain $\Omega$ is shown in figure \ref{fig:omega_var}. Despite that $\mvec\beta^{\text{OLS}}\notin\Omega$, the Lasso regularisation path must have some points in $\Omega$, since it is a continuous path that ends at $\mvec 0$.

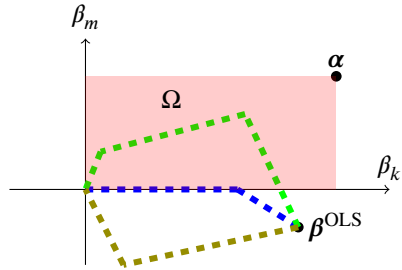
\begin{figure}[H]
    \centering
    \begin{tikzpicture}[scale=1.0]
        \draw[->] (-1, 0) -- (4, 0) node[above] {$\beta_k$};
        \draw[->] (0, -1) -- (0, 2) node[above] {$\beta_m$};
        \fill (3.3,1.5)  circle[radius=2pt] node[above] {$\mvec\alpha$};
        \fill (2.8,-.5)  circle[radius=2pt] node[right] {$\mvec\beta^{\text{OLS}}$};
        
        \draw[blue,dashed,line width=0.8mm,] (2.8,-.5) -- (2,0);
        \draw[blue,dashed,line width=0.8mm,] (2,0) -- (0,0);
        \draw[green,dashed,line width=0.8mm,] (2.8,-.5) -- (2.1,1);
        \draw[green,dashed,line width=0.8mm,] (2.1,1) -- (.2,.5);
        \draw[green,dashed,line width=0.8mm,] (.2,.5) -- (0,0);
        \draw[olive,dashed,line width=0.8mm,] (2.8,-.5) -- (.5,-1);
        \draw[olive,dashed,line width=0.8mm,] (.5,-1) -- (0,0);
        
        \draw[fill=red,fill opacity=0.2, draw=none] (0,0) rectangle (3.3, 1.5);
        \node[text width=0cm] at (1,1.2) {\red{$\Omega$}};
    \end{tikzpicture}
    \caption{Example 2D projection of the domain $\Omega$. The dashed lines represent three possible regularisation paths.}
    \label{fig:omega_var}
\end{figure}
\end{remark}

\section{Polynomial Chaos Expansion}
\label{app:PCE}

The PCE method is a surrogate based approach to UQ and sensitivity analysis. With this method a  surrogate model ${\surr{n}}$ is trained, and can then be used to compute the moments and Sobol indices \cite{saltelli_sensitivity_1995} of ${\surr{n}}$. The surrogate model is of the type

\begin{equation}
    {\surr{n}}(\mvec{x}) = \sum_{\mvec\alpha=1}^P\beta_\alpha\Psi_{\mvec\alpha}(\mvec x),
    \label{eq:PCE}
\end{equation}

where $\Psi_{\mvec1},\Psi_{\mvec2},...,\Psi_{\mvec P}$ is a set of multivariate polynomials. These multivariate polynomials consist in the product of univariate polynomials

\begin{equation*}
    \Psi_{\mvec\alpha}(\mvec{x}) = \prod_{i=1}^d\psi_{\alpha_i}(x_i),\quad\text{with}\quad \mvec\alpha=(\alpha_1, \alpha_2,...,\alpha_d), \quad \text{and}\quad \alpha_i>0,
\end{equation*}

where $\psi_{\alpha_i}(x_i)$ is a univariate polynomial of degree $\alpha_i$. The univariate polynomials are chosen such that they form an orthogonal basis with respect to the input distribution, such that we have

\begin{equation*}
    \int_{-\infty}^\infty\psi_m(x)\psi_n(x)\phi(x)\,dx = 0 \iff n\neq m,
\end{equation*}

where $\phi(x)$ is the PDF of the input distribution. As an example, for a Gaussian input distribution one would choose the Hermite polynomials as the basis, and for a uniform input distribution we would use the Legendre polynomials. Given a family of orthogonal polynomials, if $f(\mvec x) \in L^2\big(\mathbb{R}^d,\phi(\mvec x)d\mvec{x}\big)$ the following expansion

\begin{equation}
    f(\mvec{x}) = \sum_{\mvec{\alpha}=1}^\infty\beta_\alpha\Psi_{\mvec\alpha}(\mvec{x})
    \label{eq:PCE_infty}
\end{equation}

is convergent in the $L2$ sense \cite{cameron_orthogonal_1947}. However, expression \eqref{eq:PCE_infty} needs to be truncated to have polynomials of a maximum degree $p$. Given a maximum degree $p$, and an input dimension $d$, the number of polynomials that we can build is

\begin{equation}
    P=\begin{pmatrix}d+p\\p\\\end{pmatrix} = \frac{(d+p)!}{p!d!}.
\end{equation}

Hence we end up with expression \eqref{eq:PCE}, that needs to have its parameters $\beta_1,\beta_2,...,\beta_P$ fitted to a set of given data points. With this model it is then straightforward to compute the mean and variance:

\begin{equation}
    \begin{split}
        \mu_{PCE} &= \beta_1,\\
        \sigma^2_{PCE} &= \sum_{\alpha=2}^P\beta^2_i\E\left[\Psi_i^2\right].
    \end{split}
\end{equation}

In section \ref{sec:PCE_benchmark}, PCE was used with $d=8$ when comparing it to LMC, whereas the benchmark with LMC alone used $d=400$. Table \ref{tab:P_for_PCE} shows the the number of trainable parameters in a PCE model of these dimensions.

\begin{table}[H]
    \centering
    \begin{tabular}{c|c c}
         P & $d=8$ & $d=400$   \\
         \hline
         $p = 2$ & 45 & 80601\\
         $p = 3$ & 165 & $\sim 10^7$\\
         $p = 4$ & 495 & $\sim 10^9$\\
         $p = 5$ & 1287 & $\sim 10^{11}$\\
    \end{tabular}
    \caption{Number of multivariate polynomials $P$ that can be generated with a given orthogonal basis, for an input dimension $d$ and maximum degree $p$.}
    \label{tab:P_for_PCE}
\end{table}

\end{document}